\documentclass[11pt]{article}
\usepackage{natbib}
\usepackage{amsfonts}
\usepackage{amsmath}
\usepackage{amssymb}
\usepackage{epsfig}
\usepackage{float}
\usepackage[onehalfspacing]{setspace}
\usepackage{graphicx}
\usepackage{caption}
\usepackage{subcaption}
\usepackage{geometry,enumerate,bbm}
\usepackage{lscape}
\usepackage[flushleft]{threeparttable}
\usepackage{ulem}
\usepackage{scrextend}
\usepackage{placeins}

\RequirePackage[colorlinks,citecolor=blue,urlcolor=blue,linkcolor=blue]{hyperref}

\newcommand\redsout{\bgroup\markoverwith{\textcolor{red}{\rule[0.5ex]{2pt}{2.0pt}}}\ULon}
\newcommand*\boskip[1]{}

\newcommand*\bocaption[1]{\newline  {#1}}

\input{tcilatex}
\begin{document}

\author{ Bo E. Honor\'{e}\thanks{%
Princeton University, USA. Email: {honore@princeton.edu} (Corresponding
Author)} \and Luojia Hu\thanks{%
Federal Reserve Bank of Chicago, USA. Email: {lhu@frbchi.org}} \and %
Ekaterini Kyriazidou\thanks{%
New York University Abu Dhabi, UAE. Email: ak7482@nyu.edu} \and Martin
Weidner\thanks{%
University of Oxford, UK. Email: {martin.weidner@economics.ox.ac.uk}}}
\title{Simultaneity in Binary Outcome Models with an Application to
Employment for Couples\thanks{
The paper was prepared for the \textit{Special Issue of Empirical Economics
in Honor of Peter Schmidt}. The research was supported by the Gregory C.
Chow Econometric Research Program at Princeton University, by the National
Science Foundation (Grant Number SES-1530741) and by the European Research
Council through the grant ERC-2018-CoG-819086-PANEDA. The opinions expressed
here are those of the authors and not necessarily those of the Federal
Reserve Bank of Chicago or the Federal Reserve System. The data, the Matlab
programs used for the estimation and the Mathematica program that derives
the moment conditions in the Appendix will be made available online upon
publication of the paper.}}
\date{March 2023}
\maketitle

\begin{abstract}
\singlespacing
\noindent Two of Peter Schmidt's many contributions to econometrics have
been to introduce a simultaneous logit model for bivariate binary outcomes
and to study estimation of dynamic linear fixed effects panel data models
using short panels. In this paper, we study a dynamic panel data version of
the bivariate model introduced in \cite{SchmidtStrauss1975} that allows for
lagged dependent variables and fixed effects as in \cite{Ahn1995}. We
combine a conditional likelihood approach with a method of moments approach
to obtain an estimation strategy for the resulting model. We apply this
estimation strategy to a simple model for the intra-household relationship
in employment. Our main conclusion is that the within-household dependence
in employment differs significantly by the ethnicity composition of the
couple even after one allows for unobserved household specific heterogeneity.

\bigskip
\end{abstract}



\pagenumbering{gobble}

\newpage

\section*{Compliance with Ethical Standards}

\textbf{Conflict of interest:} The authors declare that they have no
conflict of interests.

\medskip

\noindent \textbf{Ethical approval:} This article does not contain any
studies with human participants or animals performed by any of the authors.

\newpage

\pagenumbering{arabic}

\section{Introduction}

A large recent literature has been concerned with econometric models in
which binary outcomes interact with each other. The papers by \cite%
{bresnahanreiss1991} and \cite{tamer2003} are early examples of this. In
those papers, the dependence is due to strategic interactions between
economic agents. This literature was predated by \cite{SchmidtStrauss1975}
who proposed a reduced form statistical model that has the feature that the
conditional distribution of each binary variable depends on the outcome of
the other.

At the same time, a large econometric literature has been concerned with
estimation of linear panel data models with fixed effects and lagged
dependent variables. This literature dates back to \cite{Nickell81} and \cite%
{AndHsi82}. The paper by \cite{Ahn1995} is an important contribution to this
literature.

This paper combines insights from these literatures by illustrating how the
simultaneous binary outcome model in \cite{SchmidtStrauss1975} can be
modified to allow for panel data with individual specific fixed effects and
lagged dependent variables. The main contribution of the paper is to develop
a toolbox of estimation procedures that can be used to estimate the
resulting models.

Methodologically, the paper fits into the literature that is concerned with
estimation of standard nonlinear panel data models with fixed effects using
short panels. This literature has a long history in econometrics. The main
problem to be solved is that treating the fixed effects as parameters to be
estimated will typically lead to inconsistent estimation of all the model
parameters. The literature has developed a number of methods to deal with
this. One approach for parametric models is to try to construct a
non-trivial sufficient statistic for the fixed effect. If such a sufficient
statistic exists, then conditional maximum likelihood (conditional on this
sufficient statistic) can typically be used to estimate the parameters of
the model. This approach was, for example, taken by \cite{Rasch60} and \cite%
{HausmanHallGriliches84} for the logit model and the Poisson regression
model, respectively. \cite{Manski87} proposed a conditional maximum score
estimator for the semiparametric binary response model with fixed effects,
which can be thought of as a generalization of the conditional maximum
likelihood approach. \cite{HonKyr00a} adapted both the conditional maximum
likelihood and the conditional maximum score methods to binary outcome
models with lagged dependent variables and fixed effects. A second strand of
the literature has studied specific semiparametric models and has been able
to find moment conditions which do not depend on the fixed effects, and
which can therefore be used to estimate the model parameters via generalized
method of moments. See for example, \cite{Honore92}, \cite{Chamberlain1992},
\cite{Kyriazidou97}, \cite{Wooldridge1997} , \cite{Kyriazidou01} and \cite%
{Hu2002}. More recently,  \cite{johnson2004identification},  \cite%
{kitazawa2013exploration}, \cite{HonoreWeidner2020} and \cite%
{HonoreMurisWeidner2021} and \cite{davezies2022fixed} have derived moment
conditions for parametric logit-type models with fixed effects, for which
the conditional likelihood approach cannot be applied.

In this paper, we study estimation of a dynamic fixed effects panel data
version of the Schmidt-Strauss model. It turns out that although the
conditional likelihood approach can be applied to identify and estimate some
of the parameters of the model, it does not identify the key parameter that
captures the dependence between the binary outcomes. On the other hand, it
turns out that one can construct moment conditions that do depend on this
parameter, which can therefore be estimated by generalized method of moments.

As an empirical illustration of the models and methods studied in this
paper, we investigate the joint determination of husbands' and wives'
employment. In this context, it is natural to allow for the possibility that
the outcome for each spouse is related to the outcome of the other, which
makes it natural to consider the Schmidt-Strauss framework. The specific
empirical question is how the parameter that captures the dependence between
outcomes for husbands and wives differs by the ethnicity of the couple, and
whether it varies over time. Since there is likely persistence in
employment, and that some of this persistence might be due to heterogeneity
as opposed to true state dependence, it is therefore natural to study this
question using dynamic panel data versions of the model proposed by \cite%
{SchmidtStrauss1975}.

The paper is organized as follows: In Section 2, we present the \cite%
{SchmidtStrauss1975} model. In Section 3, we discuss the data. Section 4
presents simple evidence for the intra-household dependence in couples%
’ employment by ethnicity. Section 5 develops and discusses a
conditional likelihood approach for estimating a version of the Schmidt and
Strauss model that incorporates lagged dependent variables as well as fixed
effects. Section 6 discusses how the method of moments approach of \cite%
{HonoreWeidner2020} can be used to identify the dependence parameter. In
Section 7, we compare the fixed effects approach to a correlated random
effects approach in the spirit of \cite{Wooldridge2005}. Section 8
concludes. The Appendix provides moment conditions for a special case of the
model in Section 6.

\section{The Schmidt-Strauss Model\label{SS Model}}

\cite{SchmidtStrauss1975} proposed a cross sectional simultaneous equations
logit model in which two binary variables, $y_{1,i}$ and $y_{2,i}$, for a
unit $i$ are each distributed according to a logit model conditional on the
other variable and on a set of explanatory variables%
\begin{eqnarray}
P\left( \left. y_{1,i}=1\right\vert y_{2,i},x_{1,i},x_{2,i}\right)
&=&\Lambda \left( x_{1,i}^{\prime }\beta _{1}+\rho y_{2,i}\right) ,
\label{EQ: SS First explained} \\
P\left( \left. y_{2,i}=1\right\vert y_{1,i},x_{1,i},x_{2,i}\right)
&=&\Lambda \left( x_{2,i}^{\prime }\beta _{2}+\rho y_{1,i}\right) .  \notag
\end{eqnarray}%
Here $x_{1,i}$ and $x_{2,i}$ are vectors of explanatory variables, $\beta_1$%
, $\beta_2$ and $\rho$ are parameters to be estimated, and $\Lambda \left(
\cdot \right) $ is the logistic cumulative distribution function. The
parameter $\rho$ captures the dependence between $y_{1,i}$ and $y_{2,i}$.
\cite{SchmidtStrauss1975} show that this model cannot be generalized to
allow for different values for $\rho $ in the distribution of $y_{1,i}$
given $y_{2,i}$ and in the distribution of $y_{2,i}$ given $y_{1,i}$. In
this sense, $\rho$ resembles the covariance between two random variables.
When the parameter $\rho $ is positive (negative), the probability that $%
y_{1,i}$ equals one is higher (lower) conditional on $y_{2,i}$ being one
than conditional on $y_{2,i}$ being zero. The same holds for the probability
that $y_{2,i}$ is one conditional on $y_{1,i}$. Holding the explanatory
variables fixed, a positive (negative) $\rho $ therefore corresponds to a
positive (negative) statistical association between $y_{1,i}$ and $y_{2,i}$.

The simultaneous logit model of \cite{SchmidtStrauss1975} has been applied
in a variety of cross sectional studies and in various fields such as labor
economics (for example, by \cite{lehrer1985determinants}  to study the
determinant of different aspects of a chosen occupation), urban economics
(for example, by \cite{boehm1981tenure} to study the effects of various
variables on the choice to own or rent and on expected future mobility),
health economics (for example, by \cite{akin1981demand} to study the use of
different kinds of health services, and by \cite{wang2007perceived} to study
the need for health insurance on one hand and actual purchase of health
insurance on the other), transportation (for example, by \cite%
{ye2007exploration} to study the relationship between mode of transportation
and trip chaining), political economy (for example, by \cite{kau1982general}
to study the interactions between congressional voting, campaign
contributions and electorial margins), and demography (for example, by \cite%
{koo1983interrelationships} to study the relationship between the
probability of dissolving a marriage and of having a child).

\begin{figure}[h]
\caption{The Relationship between $\protect\rho$ and the Correlation
Coefficient}
\label{FIGURE: Interpretation of rho}
\begin{center}
\includegraphics[width = 0.5\textwidth]{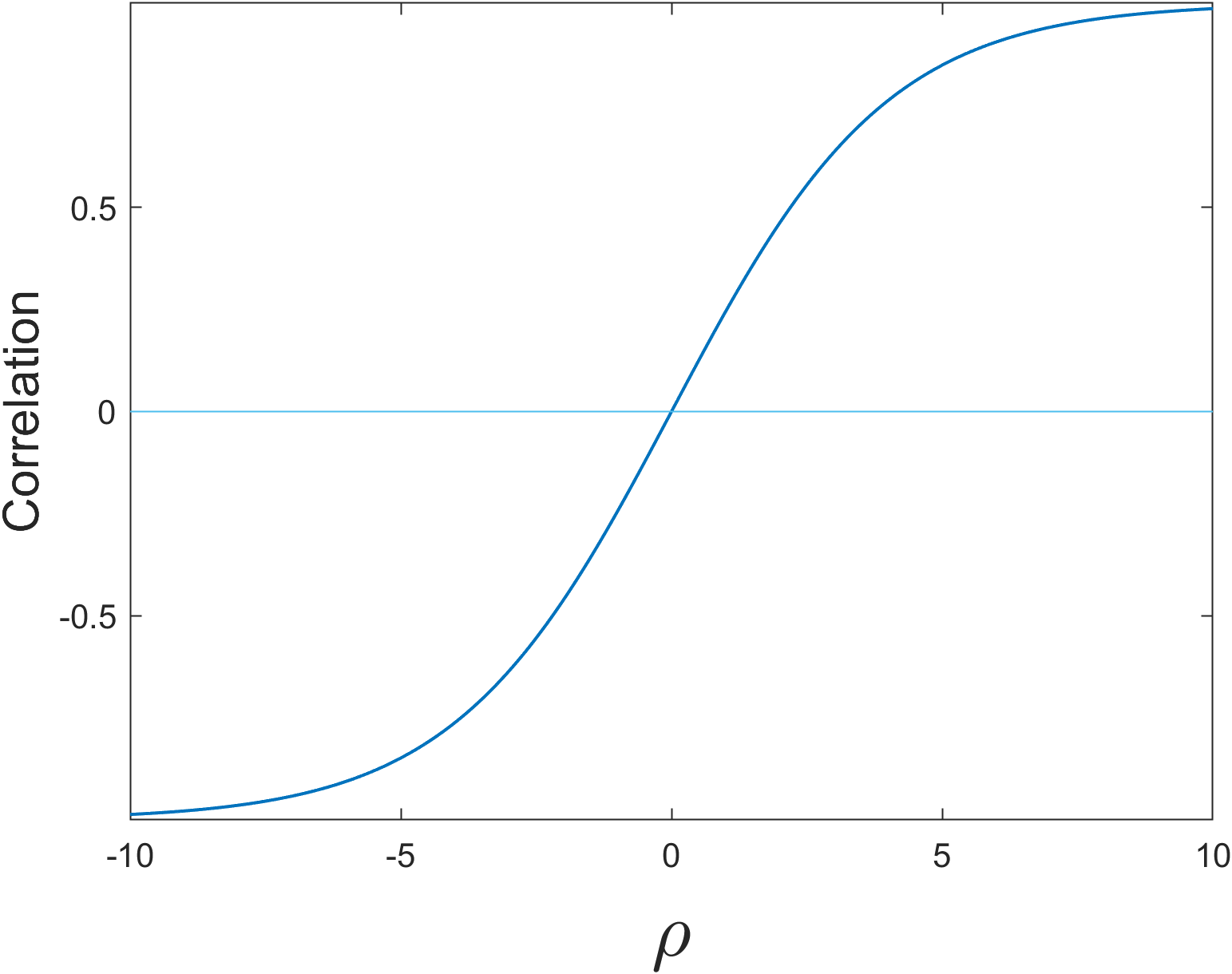}
\end{center}
\par
{\footnotesize \singlespacing  The figure shows the correlation between two
Bernoulli random variables from the model in equation (\ref{EQ: SS
Probability}), each with probability of success equal to $\frac{1}{2}$ as a
function of the parameter $\rho$. }
\end{figure}

The conditional probabilities in equation (\ref{EQ: SS First explained})
emerge from a statistical model in which $y_{1,i}$ and $y_{2,i}$ have the
joint probability distribution
\begin{align}
& P\left( \left. y_{1,i}=c_{1},y_{2,i}=c_{2}\right\vert
x_{1,i},x_{2,i}\right)  \label{EQ: SS Probability} \\
& =\frac{\exp \left( c_{1}x_{1,i}^{\prime }\beta _{1}+c_{2}x_{2,i}^{\prime
}\beta _{2}+c_{1}c_{2}\rho \right) }{1+\exp \left( x_{1,i}^{\prime }\beta
_{1}\right) +\exp (x_{2,i}^{\prime }\beta _{2})+\exp \left( x_{1,i}^{\prime
}\beta _{1}+x_{2,i}^{\prime }\beta _{2}+\rho \right) }.  \notag
\end{align}%
Another way to see that $\rho $ measures the dependence between $y_{1,i}$
and $y_{2,i}$ in equation (\ref{EQ: SS Probability}), is to note that
\begin{eqnarray}
\rho &=&\log \left( P\left( \left. y_{1,i}=1,y_{2,i}=1\right\vert
x_{1,i},x_{2,i}\right) \right) +\log \left( P\left( \left.
y_{1,i}=0,y_{2,i}=0\right\vert x_{1,i},x_{2,i}\right) \right)
\label{EQ: Rho} \\
&&-\log \left( P\left( \left. y_{1,i}=0,y_{2,i}=1\right\vert
x_{1,i},x_{2,i}\right) \right) -\log \left( P\left( \left.
y_{1,i}=1,y_{2,i}=0\right\vert x_{1,i},x_{2,i}\right) \right) .  \notag
\end{eqnarray}%
Therefore, $\log \left( P\left( \left.
y_{1,i}=c_{1},y_{2,i}=c_{2}\right\vert x_{1,i},x_{2,i}\right) \right) $ is
supermodular or submodular depending on whether $\rho >0$ or $\rho <0$. To
understand how the magnitude of $\rho$, as opposed to its sign, translates
into other measures of dependence, one can consider the following thought
experiment: Suppose that, for a given $\rho $, $\beta _{1}$ and $\beta _{2}$
above are chosen such that $y_{1,i}$ and $y_{2,i}$ are Bernoulli, each with%
\footnote{%
The reason why we focus on the case where the two probabilities
are
equal is that different values of the probabilities will bound the
correlation away from $-1$ or $1$.} probability of success equal to $0.5$.
The correlation between $y_{1,i}$ and $y_{2,i}$ then  relates to $\rho $ as
depicted in Figure \ref{FIGURE: Interpretation of rho}.

Below, we apply the model of \cite{SchmidtStrauss1975} (and its panel data
extensions) to an empirical study of husbands' and wives' employment status.
In this context, $i$ denotes the identity of the household, and $y_{1,i}$
and $y_{2,i}$ will denote the employment status of the wife and the husband,
respectively. The next section introduces the data.

\section{Data}

For the analysis in this paper, we use the Current Population Survey (CPS)
Basic Monthly micro data from the 40 years between January of 1982 and
December of 2021. The data are sourced from https://www.ipums.org/\ (\cite%
{ipums2021}). The monthly CPS has a panel design. Households are interviewed
for four consecutive months, then not interviewed for eight months, and
finally interviewed for four more consecutive months. We identify households
with one head of household and one married or unmarried partner (of the
head). The data consist of these heads and partners provided that they are
of different sex and are both between the age of 25 and 65 (inclusive).%
\footnote{%
We further clean the data by eliminating individuals with missing or
logically inconsistent age increases from one period to the next or
inconsistent sex or race over time.} Below, we sometimes refer to the
partners as husbands and wives or as spouses although they are not always
legally married. Since our ultimate goal is to investigate the dynamics of
the employment status and a number of missing observations are missing in
the last four months, we restrict the sample to the first four interview
months, and we only use households who are in the sample in all of those
four months.

We define four race/ethnicity groups: White, Black, Hispanic, and Other.
Below we interchangeably refer to these groups as \textquotedblleft
race,\textquotedblright \textquotedblleft ethnicity\textquotedblright\ or
\textquotedblleft race/ethnicity\textquotedblright . The couples are then
grouped into five groups based on the race/ethnicity of the two partners:
White, Black, Hispanic, Other, and Mixed Race. For example, White will refer
to a couple, where both spouses are White, and ``Mixed'' will refer to a
couple where the wife and husband have different ethnicity. We refer to
these groups as the \textquotedblleft ethnicity mix\textquotedblright\ (or
sometimes just the \textquotedblleft ethnicity\textquotedblright ) of the
couple.

Table \ref{TABLE: Summary Statistics} presents summary statistics for the
variables used in this paper. The first is a dummy variable for working
defined as the employment status being ``At work''. The remaining variables
are age in years, a dummy variable for the presence of children under the
age of 5, a dummy variable for any children, and dummy variables for three
education levels: high school or less, some college and college degree or
more. Note that we report the number of individuals. Since this is a
balanced panel with four time periods, the number of observations is larger
than the number of individuals by a factor of four.

\begin{table}[h!]
\caption{Summary Statistics by Household Ethnicity}
\label{TABLE: Summary Statistics}
\begin{center}
\begin{tabular}{lrrrrrr}
& \multicolumn{6}{c}{Women} \\
& All & Whites & Blacks & Hispanics & Other & Mixed \\
Working & 0.64 & 0.65 & 0.67 & 0.52 & 0.62 & 0.67 \\
Age & 43.35 & 43.80 & 43.41 & 40.61 & 41.92 & 41.26 \\
Kids $<$ 5 & 0.19 & 0.18 & 0.18 & 0.28 & 0.25 & 0.23 \\
Kids & 0.65 & 0.63 & 0.69 & 0.81 & 0.77 & 0.65 \\
HS or Less & 0.50 & 0.49 & 0.53 & 0.73 & 0.41 & 0.39 \\
Some College & 0.23 & 0.24 & 0.26 & 0.16 & 0.18 & 0.28 \\
College+ & 0.27 & 0.28 & 0.21 & 0.10 & 0.42 & 0.33 \\
No. Individuals & 1,002,489 & 783,312 & 54,342 & 63,999 & 39,765 & 61,071 \\
\hline
&  &  &  &  &  &  \\
& \multicolumn{6}{c}{Men} \\
& All & Whites & Blacks & Hispanics & Other & Mixed \\
Working & 0.83 & 0.84 & 0.76 & 0.83 & 0.82 & 0.84 \\
Age & 45.53 & 45.93 & 45.88 & 42.81 & 44.76 & 43.52 \\
Kids $<$ 5 & 0.19 & 0.18 & 0.18 & 0.28 & 0.25 & 0.23 \\
Kids & 0.65 & 0.63 & 0.69 & 0.81 & 0.77 & 0.65 \\
HS or Less & 0.50 & 0.48 & 0.60 & 0.75 & 0.38 & 0.39 \\
Some College & 0.22 & 0.22 & 0.23 & 0.15 & 0.17 & 0.28 \\
College+ & 0.29 & 0.30 & 0.17 & 0.10 & 0.44 & 0.33 \\
No. Individuals & 1,002,489 & 783,312 & 54,342 & 63,999 & 39,765 & 61,071 \\
\hline
\end{tabular}%
\end{center}
\par
{\footnotesize \singlespacing The table shows averages by the ethnicity of
the couple for the variables used in this paper. The data are from IPUMS CPS
and cover a balanced panel of couples where each individual's age is between
25 and 65. The data cover the period between 1982 and 2021.}
\end{table}

\section{Model and Simple Evidence}

\subsection{Summary Statistics\label{SEC: Summary Statistics}}

We start by presenting summary statistics for the joint probability of
working by ethnicity. The first panel of Table \ref{TABLE: Joint Probability
of Working by Ethnicity} is for the whole sample, while the next two panels
are for the subsamples of couples without children and with children. Our
main takeaway from this table is that there is a large difference in these
probabilities across the ethnicities, with Hispanic-Hispanic couples looking
quite different from the others.

\begin{table}[h!]
\caption{Joint Probabilities of Employment by Household Ethnicity}
\label{TABLE: Joint Probability of Working by Ethnicity}%
\begin{tabular}{lccccccccccc}
&  & \multicolumn{2}{c}{White} & \multicolumn{2}{c}{Black} &
\multicolumn{2}{c}{Hispanic} & \multicolumn{2}{c}{Other} &
\multicolumn{2}{c}{Mixed} \\ \hline
&  &  &  &  &  &  &  &  &  &  &  \\
&  & \multicolumn{10}{c}{All} \\
&  & \multicolumn{2}{c}{Husband} & \multicolumn{2}{c}{Husband} &
\multicolumn{2}{c}{Husband} & \multicolumn{2}{c}{Husband} &
\multicolumn{2}{c}{Husband} \\
&  & No & Yes & No & Yes & No & Yes & No & Yes & No & Yes \\
Wife & No & $0.087 $ & $0.260 $ & $0.114 $ & $0.217 $ & $0.096 $ & $0.388 $
& $0.088 $ & $0.294 $ & $0.074 $ & $0.258 $ \\
& Yes & $0.076 $ & $0.578 $ & $0.129 $ & $0.540 $ & $0.071 $ & $0.444 $ & $%
0.087 $ & $0.531 $ & $0.091 $ & $0.577 $ \\
&  &  &  &  &  &  &  &  &  &  &  \\
&  & \multicolumn{10}{c}{Without Children} \\
&  & \multicolumn{2}{c}{Husband} & \multicolumn{2}{c}{Husband} &
\multicolumn{2}{c}{Husband} & \multicolumn{2}{c}{Husband} &
\multicolumn{2}{c}{Husband} \\
&  & No & Yes & No & Yes & No & Yes & No & Yes & No & Yes \\
Wife & No & $0.096 $ & $0.231 $ & $0.124 $ & $0.202 $ & $0.106 $ & $0.342 $
& $0.093 $ & $0.255 $ & $0.081 $ & $0.226 $ \\
& Yes & $0.084 $ & $0.589 $ & $0.138 $ & $0.537 $ & $0.080 $ & $0.471 $ & $%
0.096 $ & $0.557 $ & $0.100 $ & $0.593 $ \\
&  &  &  &  &  &  &  &  &  &  &  \\
&  & \multicolumn{10}{c}{With Children} \\
&  & \multicolumn{2}{c}{Husband} & \multicolumn{2}{c}{Husband} &
\multicolumn{2}{c}{Husband} & \multicolumn{2}{c}{Husband} &
\multicolumn{2}{c}{Husband} \\
&  & No & Yes & No & Yes & No & Yes & No & Yes & No & Yes \\
Wife & No & $0.044 $ & $0.389 $ & $0.072 $ & $0.283 $ & $0.071 $ & $0.508 $
& $0.073 $ & $0.412 $ & $0.053 $ & $0.364 $ \\
& Yes & $0.039 $ & $0.528 $ & $0.090 $ & $0.555 $ & $0.048 $ & $0.373 $ & $%
0.062 $ & $0.453 $ & $0.059 $ & $0.523 $ \\ \hline
\end{tabular}
\bocaption{\footnotesize  \singlespacing The table shows the fraction of couples in each group that
report each combination of working and not working. The data are from IPUMS CPS and cover a balanced panel of couples
where each individual's age is between 25 and 65. The data cover the period between 1982 and 2021.}
\end{table}

Table \ref{TABLE: Joint Probability of Working by Ethnicity}\ aggregates the
data for all years. In Figure \ref{FIGURE: Probability Distribution of
Working} we plot the joint probability of working over time for each
ethnicity. These are depicted in the four leftmost plots. The two plots to
the right are the marginal probabilities of working for the husbands and
wives. Again, the main takeaway is that there are interesting differences
across ethnicities, with Hispanics and, to a lesser extent, Blacks standing
out. In terms of the evolution of the probabilities over time, the most
distinct feature is the increase in the employment of women in the first
part of the sample. This is seen in the marginal probabilities as well as
the joint probabilities. It is also interesting that the 2008 recession had
a large impact on the employment of men, but almost no effect for the women.

\begin{figure}[h]
\caption{Probability Distributions of Employment over Time by Household
Ethnicity}
\label{FIGURE: Probability Distribution of Working}
\begin{center}
\includegraphics[width = 0.9\textwidth]{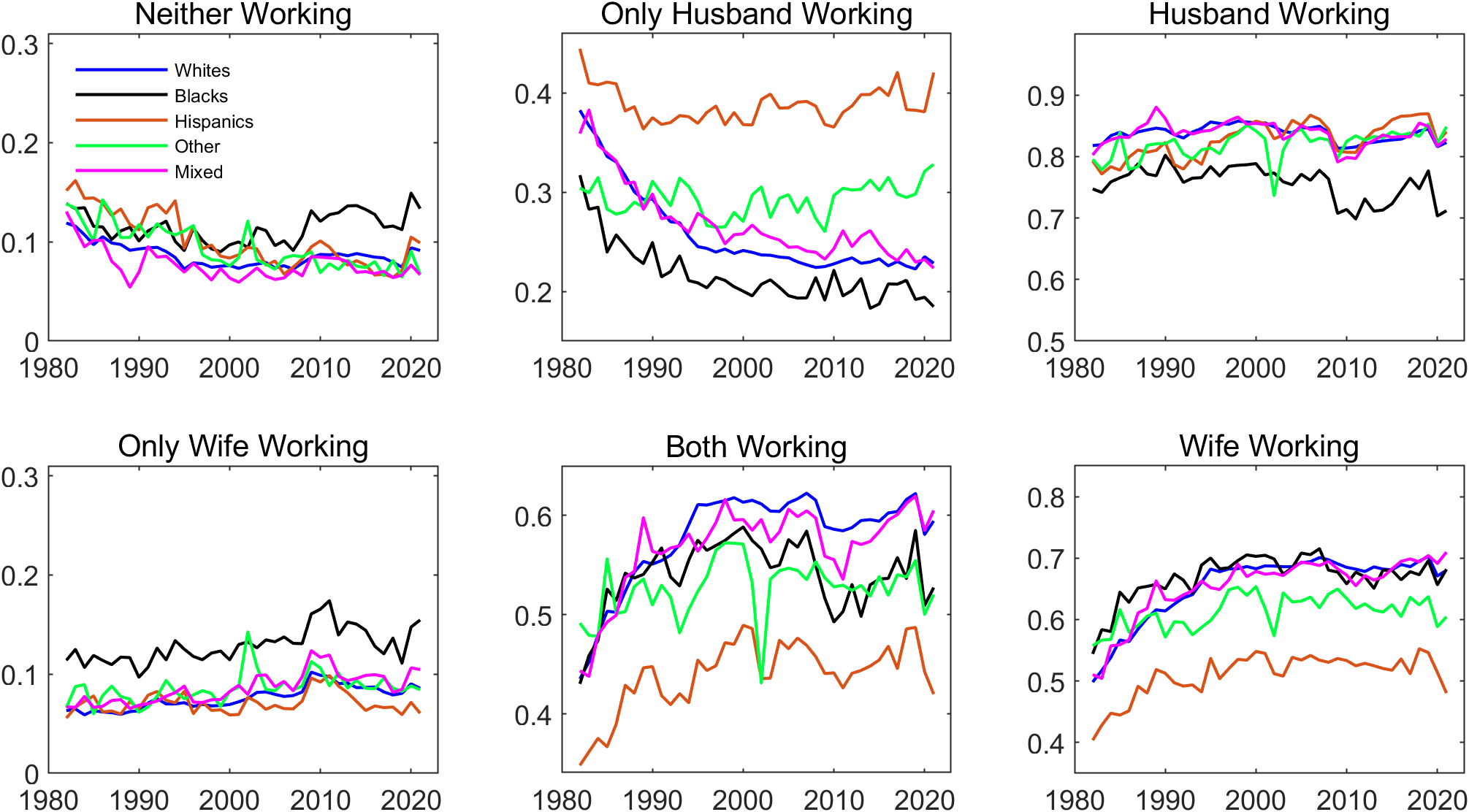}
\end{center}
\par
{\footnotesize \singlespacing The data are from IPUMS CPS and cover a
balanced panel of couples where each individual's age is between 25 and 65.
The data cover the period between 1982 and 2021.}
\end{figure}

The left panel of Figure \ref{FIGURE: Within Household Correlations In
Employment Over Time} displays the correlation between the spouses'
employment over time. The reported correlation is a five year centered
moving average. The correlation is always positive for all of the
ethnicities. For Blacks and Whites, it remained more or less stable over
time, while it decreased dramatically for the other groups, especially for
Hispanics and for Others. It is difficult to compare correlations of
different pairs of binary variables when the marginal probabilities differ
across the pairs. In the right panel of Figure \ref{FIGURE: Within Household
Correlations In Employment Over Time}, we therefore present the five year
centered moving average of the estimate of the parameter $\rho $ in a
Schmidt-Strauss model with no explanatory variables. Here $\hat\rho $ is
calculated by the sample analog of equation (\ref{EQ: Rho}). The estimated
trend for $\rho $ is similar to that for the correlation, although $\rho $
shows a larger difference between Whites and Blacks.

\begin{figure}[]
\caption{Evolution of Intra-Household Employment Dependence over Time by
Household Ethnicity}
\label{FIGURE: Within Household Correlations In Employment Over Time}
\begin{center}
\includegraphics[width = 0.9\textwidth]{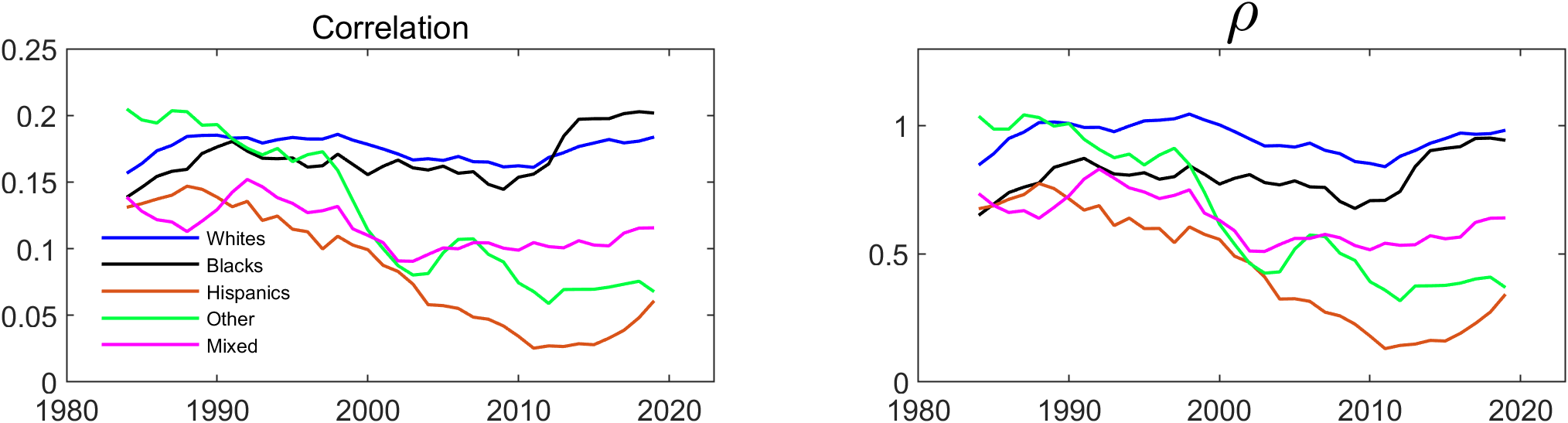}
\end{center}
\par
{\footnotesize \singlespacing  The data are from IPUMS CPS and cover a
balanced panel of couples where each individual's age is between 25 and 65.
The data cover the period between 1982 and 2021. $\rho $ is estimated by the
sample analog of equation (\ref{EQ: Rho})}
\end{figure}

\subsection{Static Cross Sectional Schmidt-Strauss Models}

It is clear from the evidence in Section \ref{SEC: Summary Statistics} that
there is a strong relationship between employment of husbands and of wives.
In this section, we document that this persists after controlling for a set
of observable characteristics. Specifically, in the first four columns of
Table \ref{TABLE: Static Cross Section}, we present the results from
estimating separate single-equation logit models for employment for
husbands and for wives as well as the results from maximum likelihood
estimation of the Schmidt-Strauss model in equation (\ref{EQ: SS Probability}%
). The explanatory variables are dummy variables for the presence of
children younger than 5, for any children, for the person's own ethnicity,
for the education categories \textquotedblleft some
college\textquotedblright\ and \textquotedblleft college and
above,\textquotedblright\ and dummy variables for the ethnicity of the
couple. The estimation also controls for year dummies, the age and the
age-squared of both the husband and the wife, as well as the interaction of
the ages. The last four columns present the results from estimating the same
models after also including the ethnicity and the education variables of the
spouse as explanatory variables.

\begin{table}[]
\caption{Estimates of Static Cross Sectional Models of Employment}
\label{TABLE: Static Cross Section}
{\footnotesize \centering
\begin{tabular*}{1.0\textwidth}{l@{\extracolsep{\fill}}rrrrrrrr}
\hline
& \multicolumn{2}{c}{Univariate Logits} & \multicolumn{2}{c}{Schmidt-Strauss}
& \multicolumn{2}{c}{Univariate Logits} & \multicolumn{2}{c}{Schmidt-Strauss}
\\
& Women & Men & Women & Men & Women & Men & Women & Men \\
Kids $<$ 5 & $-0.808$\rlap{***} & $0.015$\rlap{*} & $-0.814$\rlap{***} & $%
0.146$\rlap{***} & $-0.801$\rlap{***} & $0.009$\rlap{} & $-0.804$\rlap{***}
& $0.143$\rlap{***} \\
& ($0.006$) & ($0.008$) & ($0.006$) & ($0.009$) & ($0.006$) & ($0.009$) & ($%
0.006$) & ($0.009$) \\
Kids & $-0.183$\rlap{***} & $0.218$\rlap{***} & $-0.209$\rlap{***} & $0.253$%
\rlap{***} & $-0.180$\rlap{***} & $0.220$\rlap{***} & $-0.206$\rlap{***} & $%
0.254$\rlap{***} \\
& ($0.005$) & ($0.006$) & ($0.005$) & ($0.006$) & ($0.005$) & ($0.006$) & ($%
0.005$) & ($0.006$) \\
Black (Woman) & $-0.041$\rlap{} &  & $-0.055$\rlap{} &  & $-0.055$\rlap{} & $%
-0.114$\rlap{*} & $-0.045$\rlap{} & $-0.107$\rlap{*} \\
& ($0.045$) &  & ($0.045$) &  & ($0.052$) & ($0.063$) & ($0.053$) & ($0.064$)
\\
Hispanic (Woman) & $-0.095$\rlap{***} &  & $-0.111$\rlap{***} &  & $-0.132$%
\rlap{***} & $-0.003$\rlap{} & $-0.134$\rlap{***} & $0.019$\rlap{} \\
& ($0.020$) &  & ($0.020$) &  & ($0.039$) & ($0.047$) & ($0.039$) & ($0.048$)
\\
Other (Woman) & $-0.156$\rlap{***} &  & $-0.171$\rlap{***} &  & $-0.199$%
\rlap{***} & $-0.078$\rlap{*} & $-0.195$\rlap{***} & $-0.047$\rlap{} \\
& ($0.020$) &  & ($0.020$) &  & ($0.038$) & ($0.046$) & ($0.039$) & ($0.047$)
\\
Some College (Woman) & $0.381$\rlap{***} &  & $0.365$\rlap{***} &  & $0.395$%
\rlap{***} & $0.188$\rlap{***} & $0.381$\rlap{***} & $0.124$\rlap{***} \\
& ($0.005$) &  & ($0.005$) &  & ($0.005$) & ($0.007$) & ($0.005$) & ($0.007$)
\\
College+ (Woman) & $0.571$\rlap{***} &  & $0.537$\rlap{***} &  & $0.699$%
\rlap{***} & $0.228$\rlap{***} & $0.687$\rlap{***} & $0.116$\rlap{***} \\
& ($0.005$) &  & ($0.005$) &  & ($0.006$) & ($0.008$) & ($0.006$) & ($0.008$)
\\
Black (Man) &  & $-0.376$\rlap{***} &  & $-0.392$\rlap{***} & $0.109$%
\rlap{***} & $-0.409$\rlap{***} & $0.153$\rlap{***} & $-0.433$\rlap{***} \\
&  & ($0.033$) &  & ($0.033$) & ($0.039$) & ($0.046$) & ($0.040$) & ($0.047$)
\\
Hispanic (Man) &  & $0.003$\rlap{} &  & $-0.018$\rlap{} & $-0.062$\rlap{} & $%
-0.039$\rlap{} & $-0.060$\rlap{} & $-0.030$\rlap{} \\
&  & ($0.025$) &  & ($0.026$) & ($0.039$) & ($0.048$) & ($0.040$) & ($0.049$)
\\
Other (Man) &  & $-0.192$\rlap{***} &  & $-0.214$\rlap{***} & $-0.109$%
\rlap{***} & $-0.239$\rlap{***} & $-0.086$\rlap{**} & $-0.225$\rlap{***} \\
&  & ($0.028$) &  & ($0.028$) & ($0.040$) & ($0.048$) & ($0.041$) & ($0.049$)
\\
Some College (Man) &  & $0.318$\rlap{***} &  & $0.298$\rlap{***} & $0.104$%
\rlap{***} & $0.259$\rlap{***} & $0.079$\rlap{***} & $0.247$\rlap{***} \\
&  & ($0.006$) &  & ($0.006$) & ($0.005$) & ($0.007$) & ($0.005$) & ($0.007$)
\\
College+ (Man) &  & $0.742$\rlap{***} &  & $0.699$\rlap{***} & $-0.219$%
\rlap{***} & $0.631$\rlap{***} & $-0.281$\rlap{***} & $0.676$\rlap{***} \\
&  & ($0.006$) &  & ($0.006$) & ($0.006$) & ($0.007$) & ($0.006$) & ($0.007$)
\\
Black Household & $0.141$\rlap{***} & $-0.134$\rlap{***} & $0.220$\rlap{***}
& $-0.144$\rlap{***} & $0.028$\rlap{} & $0.013$\rlap{} & $0.033$\rlap{} & $%
0.007$\rlap{} \\
& ($0.045$) & ($0.035$) & ($0.046$) & ($0.035$) & ($0.077$) & ($0.093$) & ($%
0.078$) & ($0.095$) \\
Hispanic Household & $-0.419$\rlap{***} & $-0.147$\rlap{***} & $-0.392$%
\rlap{***} & $-0.028$\rlap{} & $-0.331$\rlap{***} & $-0.065$\rlap{} & $%
-0.332 $\rlap{***} & $-0.006 $\rlap{} \\
& ($0.021$) & ($0.027$) & ($0.022$) & ($0.028$) & ($0.075$) & ($0.091$) & ($%
0.076$) & ($0.093$) \\
Other Household & $-0.092$\rlap{***} & $-0.147$\rlap{***} & $-0.051$\rlap{**}
& $-0.086$\rlap{***} & $0.079$\rlap{} & $-0.016$\rlap{} & $0.080$\rlap{} & $%
-0.027$\rlap{} \\
& ($0.022$) & ($0.030$) & ($0.022$) & ($0.031$) & ($0.074$) & ($0.090$) & ($%
0.076$) & ($0.091$) \\
Mixed Household & $0.002$\rlap{} & $-0.153$\rlap{***} & $0.029$\rlap{**} & $%
-0.134$\rlap{***} & $0.042$\rlap{} & $-0.105$\rlap{**} & $0.052$\rlap{} & $%
-0.113$\rlap{**} \\
& ($0.012$) & ($0.015$) & ($0.012$) & ($0.015$) & ($0.038$) & ($0.046$) & ($%
0.039$) & ($0.047$) \\
$\rho $ &  &  &  & $0.718$\rlap{***} &  &  &  & $0.730$\rlap{***} \\
&  &  &  & ($0.005$) &  &  &  & ($0.005$) \\ \hline
\multicolumn{9}{c}{{*}** p$<$0.01, ** p$<$0.05, * p$<$0.1} \\
&  &  &  &  &  &  &  &
\end{tabular*}
\singlespacing The dependent variable is working and the parameters are
estimated by maximum likelihood. The data are from IPUMS CPS and cover a
balanced panel of couples where each individual's age is between 25 and 65.
The data cover the period between 1982 and 2021. Coefficients on year
dummies, husband's and wife's age, their interaction and their squares are
not reported. Standard errors are clustered at the household level.}
\end{table}

The estimates of $\rho $ in Table \ref{TABLE: Static Cross Section} clearly
suggest that there is positive association between the employment of
husbands and wives after controlling for observed characteristics. In order
to investigate whether this association varies systematically across
ethnicities, we re-estimate the model in the last two columns of Table \ref%
{TABLE: Static Cross Section} separately for each ethnicity. In Table \ref%
{TABLE: Static Cross Section By Ethnicity}, we report the estimated $\rho $%
's. The most striking finding is that the estimated $\rho $ for Whites is
much larger than for other ethnicities, while the estimate for Hispanics is
the lowest. This is also reflected in counterfactual marginal effects.
Specifically, for each ethnicity, we calculate the average probabilities
implied by the model that a wife works conditional on whether her husband
works or not. The difference in these average probabilities is 18 percentage
points for Whites, 8 for Hispanics, and between 11 and 14 for each of the
other three groups. The corresponding counterfactual marginal effects for
husbands are 10 percentage points for Whites, 4 for Hispanics, and between 6
and 9 percentage points for the other groups. This ordering is consistent
with that found in Figure \ref{FIGURE: Within Household Correlations In
Employment Over Time}.

\begin{table}[]
\caption{Estimates of $\protect\rho$ in the Static Cross Sectional
Schmidt-Strauss Model by Household Ethnicity}
\label{TABLE: Static Cross Section By Ethnicity}
\
\par
\begin{center}
{\footnotesize
\begin{tabular}{lrrrrr}
\hline
& White & Black & Hispanic & Other & Mixed \\
$\rho $ & $0.814$\rlap{***} & $0.540$\rlap{***} & $0.356$\rlap{***} & $0.604$%
\rlap{***} & $0.506$\rlap{***} \\
& ($0.006$) & ($0.019$) & ($0.019$) & ($0.025$) & ($0.020$) \\ \hline
\multicolumn{6}{c}{{*}** p$<$0.01, ** p$<$0.05, * p$<$0.1}%
\end{tabular}
}
\end{center}
\par
{\footnotesize \singlespacing The dependent variable is working and the
parameters are estimated by maximum likelihood using the same specification
as in Table \ref{TABLE: Static Cross Section}. The data are from IPUMS CPS
and cover a balanced panel of couples where each individual's age is between
25 and 65. The data cover the period between 1982 and 2021. Standard errors
are clustered at the household level.}
\end{table}

\begin{figure}[]
\caption{Evolution of $\protect\rho$ over Time by Household Ethnicity}
\label{FIGURE: Evolution of CS Rho's over Time}
\begin{center}
\includegraphics[width = 0.45\textwidth]{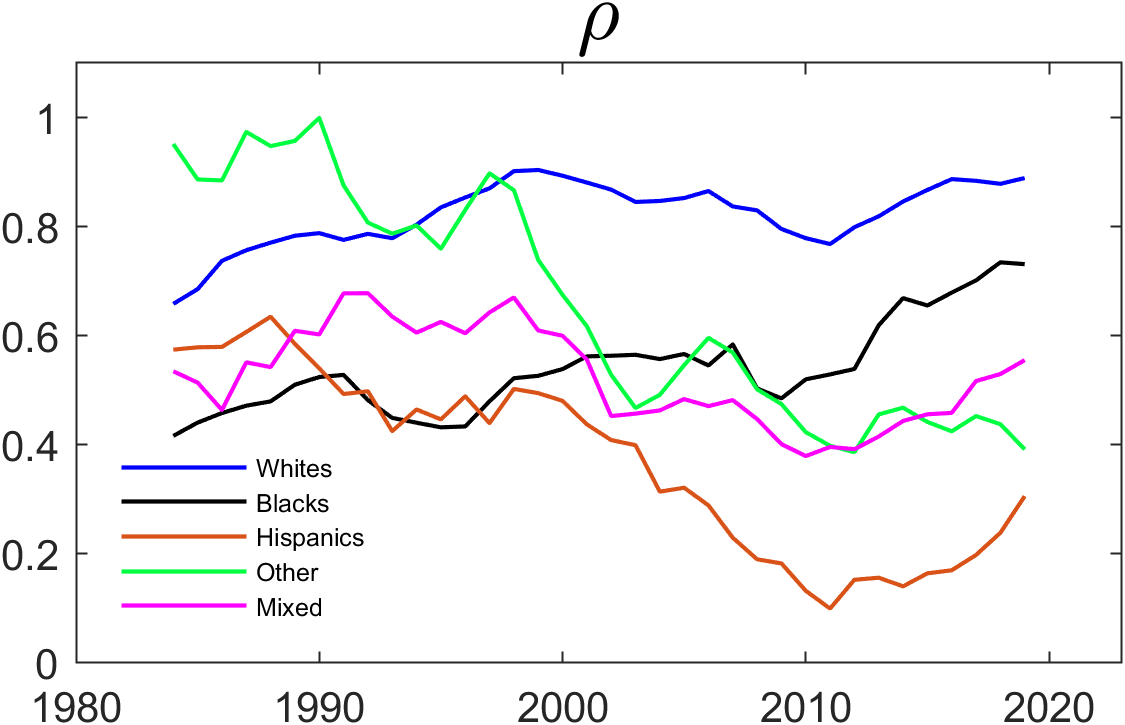}
\end{center}
\par
{\footnotesize \singlespacing The dependent variable is working and the
parameters are estimated by maximum likelihood using the same specification
as in Table \ref{TABLE: Static Cross Section}. The data are from IPUMS CPS
and cover a balanced panel of couples where each individual's age is between
25 and 65. The data cover the period between 1982 and 2021 and the
estimation is done over five year centered rolling windows.}
\end{figure}

Figure \ref{FIGURE: Within Household Correlations In Employment Over Time}
above suggested a dramatic fall in the association between the employment of
wives and husbands for households where both the wife and the husband are
Hispanic, and for households where each spouse is of ``other ethnicity''. To
investigate whether this holds after controlling for observable covariates,
we estimate the model in the last two columns of Table \ref{TABLE:
Static Cross Section} for each ethnicity and for rolling 5-year time-spans.
The estimated $\rho$ coefficients are presented in Figure \ref%
{FIGURE: Evolution of CS Rho's over Time}. Qualitatively, the pattern in
Figure \ref{FIGURE: Evolution of CS Rho's over Time} is similar to that in
Figure \ref{FIGURE: Within Household Correlations In Employment Over Time}:
The association between the employment of wives and husbands has been
falling for Hispanics and for Others, while it has been relatively stable
for White, Black and Mixed couples.

\subsection{Dynamic Panel Data Schmidt-Strauss Models}

In the Schmidt-Strauss models estimated in Table \ref{TABLE: Static Cross
Section}, the only avenue for interdependence between the employment of
wives and husbands (conditional on the observed characteristics) is through
the parameter $\rho $. If the employment of a partner actually also
depends on the lagged employment of both partners, then this will be
captured by the estimate of $\rho$.

In order to investigate the role of dynamics, we first estimate the
Schmidt-Strauss model in the last two columns of  Table \ref{TABLE:
Static Cross Section} after including an individual's own
as well as
the partner's lagged employment  as explanatory variables. Specifically, we
estimate the model
\begin{eqnarray}
&&P\left( \left. y_{1,it}=c_{1},y_{2,it}=c_{2}\right\vert \left\{
y_{1,is},y_{2,is}\right\} _{s<t},\left\{ x_{1,is}\right\} _{s=1}^{t},\left\{
x_{2,is}\right\} _{s=1}^{t}\right)  \label{EQ: SS with lags} \\
&=&\frac{\exp \left( c_{1}\left( z_{1,it}\right) +c_{2}\left(
z_{2,it}\right) +c_{1}c_{2}\rho \right) }{1+\exp \left( z_{1,it}\right)
+\exp (z_{2,it})+\exp \left( z_{1,it}+z_{2,it}+\rho \right) }  \notag
\end{eqnarray}%
for $c_{1},c_{2}\in \left\{ 0,1\right\} $, where%
\begin{eqnarray*}
z_{1,it} &=&x_{1,it}^{\prime }\beta _{1}+y_{1,it-1}\gamma
_{11}+y_{2,it-1}\gamma _{12} \\
z_{2,it} &=&x_{2,it}^{\prime }\beta _{2}+y_{1,it-1}\gamma
_{21}+y_{2,it-1}\gamma _{22}.
\end{eqnarray*}
The results are presented in Table \ref{TABLE: Including Lagged Working}.
Since the lagged values of the dependent variable are not observed in the
first time period, we do the estimation using waves two through four of our
dataset. The results in Table \ref{TABLE: Including Lagged Working} suggest
that each partner's employment depends strongly and positively on her or his
own lagged employment, and that it depends negatively on the partner's
lagged employment (after controlling for the observed covariates). In
combination, these will introduce a negative correlation in the
contemporaneous employment status, which - in turn - would lead to a
downward bias in the estimate of $\rho $ when these dynamic interactions are
not controlled for in the model. This is reflected in the higher estimate of
$\rho $ in the model that allows for lagged employment of both partners as
explanatory variables as in equation (\ref{EQ: SS
with lags}).

\begin{table}[]
\caption{Estimates of Dynamic Schmidt-Strauss Models of Employment}
\label{TABLE: Including Lagged Working}
\par
\begin{center}
{\footnotesize \
\begin{tabular*}{0.67\textwidth}{lr@{\extracolsep{\fill}}r@{\extracolsep{\fill}}r}
\hline
&  & \multicolumn{2}{c}{Schmidt-Strauss} \\
&  & Women & Men \\
Lagged Employment (Woman) &  & $4.684$\rlap{***} & $-1.668$\rlap{***} \\
&  & ($0.005$) & ($0.008$) \\
Lagged Employment (Man) &  & $-1.668$\rlap{***} & $4.343$\rlap{***} \\
&  & ($0.008$) & ($0.006$) \\
Kids $<$ 5 &  & $-0.425$\rlap{***} & $0.067$\rlap{***} \\
&  & ($0.006$) & ($0.007$) \\
Kids &  & $-0.110$\rlap{***} & $0.162$\rlap{***} \\
&  & ($0.004$) & ($0.005$) \\
Black (Woman) &  & $-0.049$\rlap{} & $-0.028$\rlap{} \\
&  & ($0.047$) & ($0.052$) \\
Hispanic (Woman) &  & $-0.097$\rlap{***} & $0.037$\rlap{} \\
&  & ($0.035$) & ($0.040$) \\
Other (Woman) &  & $-0.115$\rlap{***} & $-0.004$\rlap{} \\
&  & ($0.034$) & ($0.038$) \\
Some College (Woman) &  & $0.213$\rlap{***} & $0.077$\rlap{***} \\
&  & ($0.005$) & ($0.006$) \\
College+ (Woman) &  & $0.382$\rlap{***} & $0.077$\rlap{***} \\
&  & ($0.006$) & ($0.006$) \\
Black (Man) &  & $0.055$\rlap{} & $-0.223$\rlap{***} \\
&  & ($0.035$) & ($0.039$) \\
Hispanic (Man) &  & $-0.041$\rlap{} & $0.011$\rlap{} \\
&  & ($0.035$) & ($0.040$) \\
Other (Man) &  & $-0.054$\rlap{} & $-0.111$\rlap{***} \\
&  & ($0.036$) & ($0.041$) \\
Some College (Man) &  & $0.036$\rlap{***} & $0.151$\rlap{***} \\
&  & ($0.005$) & ($0.006$) \\
College+ (Man) &  & $-0.159$\rlap{***} & $0.393$\rlap{***} \\
&  & ($0.005$) & ($0.006$) \\
Black Household &  & $0.076$\rlap{} & $-0.077$\rlap{} \\
&  & ($0.069$) & ($0.077$) \\
Hispanic Household &  & $-0.160$\rlap{**} & $-0.068$\rlap{} \\
&  & ($0.067$) & ($0.076$) \\
Other Household &  & $0.079$\rlap{} & $-0.066$\rlap{} \\
&  & ($0.066$) & ($0.075$) \\
Mixed Household &  & $0.059$\rlap{*} & $-0.102$\rlap{***} \\
&  & ($0.034$) & ($0.039$) \\
$\rho $ &  & \multicolumn{2}{c}{$2.040$\rlap{***}} \\
&  & \multicolumn{2}{c}{($0.008$)} \\ \hline
\multicolumn{4}{c}{{*}** p$<$0.01, ** p$<$0.05, * p$<$0.1}%
\end{tabular*}
}
\end{center}
\par
{\footnotesize \singlespacing The dependent variable is working and the
parameters are estimated by maximum likelihood. The data are from IPUMS CPS
and cover a balanced panel of couples where each individual's age is between
25 and 65. The data cover the period between 1982 and 2021. Coefficients on
year dummies, husband's and wife's age, their interaction and their squares
are not reported. Standard errors are clustered at the household level.}
\end{table}

Since controlling for the lagged employment status of both partners
dramatically change the estimate of $\rho$ when we use the full sample, we
next investigate whether the same is true across ethnicities.
Specifically, we estimate the same specification as in Table \ref{TABLE: Including
Lagged Working} separately for each ethnicity group. Table \ref{TABLE:
Dynamic Cross Section By Ethnicity} reports the estimated coefficients on
the lagged employment variables as well as the estimated $\rho$. In this
specification, Hispanics and Blacks are quite similar to each other in terms
of the contemporaneous interdependence between the employment status of the
two partners (measured by $\rho$) as well as in terms of the dynamic
interdependence (measured by the $\gamma$'s).

\begin{table}[]
\caption{Estimates of Dynamic Schmidt-Strauss Models of Employment by
Household Ethnicity}
\label{TABLE: Dynamic Cross Section By Ethnicity}
\par
\begin{center}
{\footnotesize
\begin{tabular*}{1.0\textwidth}{l@{\extracolsep{\fill}}rrrrrr}
\hline
& All & Whites & Blacks & Hispanics & Other & Mixed \\
$\gamma_{11}$ & $4.684$\rlap{***} & $4.678$\rlap{***} & $4.481$\rlap{***} & $%
4.716$\rlap{***} & $4.976$\rlap{***} & $4.678$\rlap{***} \\
& ($0.005$) & ($0.006$) & ($0.023$) & ($0.021$) & ($0.029$) & ($0.022$) \\
$\gamma_{12}$ & $-1.668$\rlap{***} & $-1.759$\rlap{***} & $-1.041$\rlap{***}
& $-1.096$\rlap{***} & $-1.475$\rlap{***} & $-1.629$\rlap{***} \\
& ($0.008$) & ($0.009$) & ($0.037$) & ($0.037$) & ($0.050$) & ($0.034$) \\
$\gamma_{21}$ & $-1.668$\rlap{***} & $-1.759$\rlap{***} & $-1.061$\rlap{***}
& $-1.082$\rlap{***} & $-1.460$\rlap{***} & $-1.638$\rlap{***} \\
& ($0.008$) & ($0.009$) & ($0.037$) & ($0.036$) & ($0.050$) & ($0.034$) \\
$\gamma_{22}$ & $4.343$\rlap{***} & $4.363$\rlap{***} & $4.344$\rlap{***} & $%
4.019$\rlap{***} & $4.470$\rlap{***} & $4.359$\rlap{***} \\
& ($0.006$) & ($0.007$) & ($0.024$) & ($0.023$) & ($0.032$) & ($0.025$) \\
$\rho $ & $2.040$\rlap{***} & $2.170$\rlap{***} & $1.357$\rlap{***} & $1.262$%
\rlap{***} & $1.764$\rlap{***} & $1.868$\rlap{***} \\
& ($0.008$) & ($0.009$) & ($0.037$) & ($0.037$) & ($0.050$) & ($0.033$) \\
\hline
\multicolumn{7}{c}{{*}** p$<$0.01, ** p$<$0.05, * p$<$0.1}%
\end{tabular*}
}
\end{center}
\par
{\footnotesize \singlespacing  The dependent variable is working and the
parameters are estimated by maximum likelihood using the same specification
as in Table \ref{TABLE: Including Lagged Working}. The data are from IPUMS
CPS and cover a balanced panel of couples where each individual's age is
between 25 and 65. The data cover the period between 1982 and 2021. Standard
errors are clustered at the household level.}
\end{table}

\begin{figure}[]
\caption{Evolution of $\protect\rho$ over Time by Household Ethnicity}
\label{FIGURE: Evolution of Panel Rho's over Time}
\begin{center}
\includegraphics[width = 0.45\textwidth]{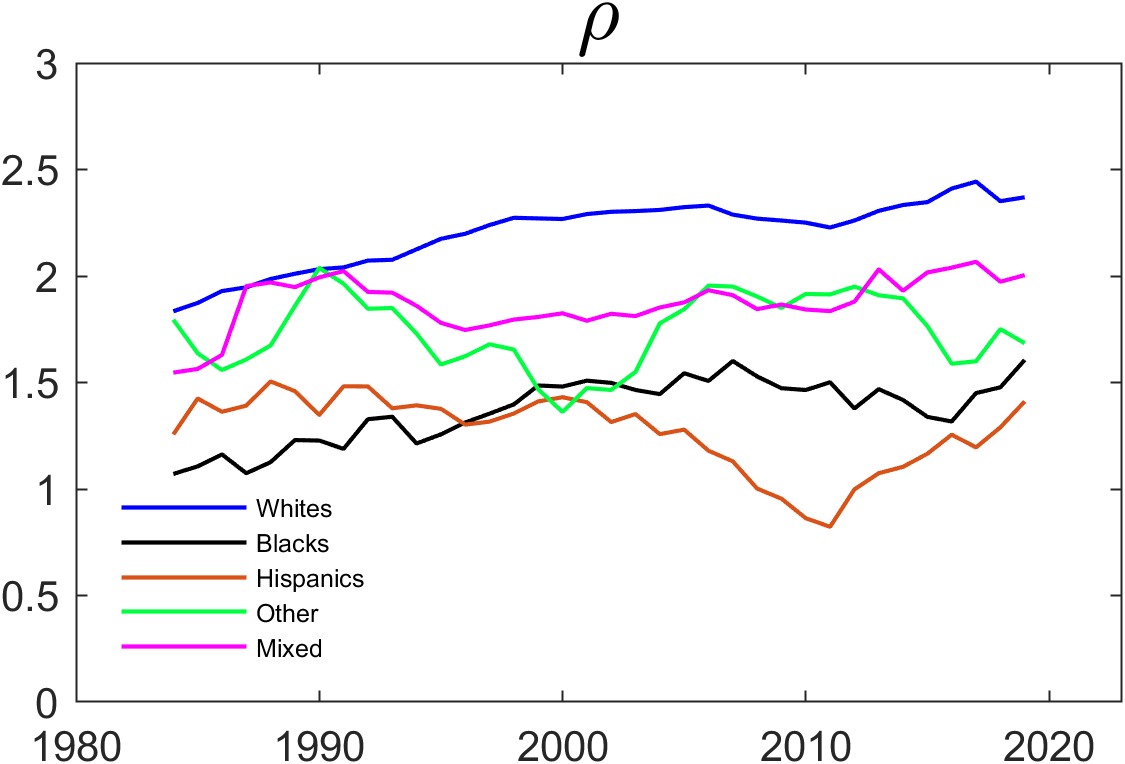}
\end{center}
\par
{\footnotesize \singlespacing  The dependent variable is working and the
parameters are estimated by maximum likelihood using the same specification
as in Table \ref{TABLE: Including Lagged Working}. The data are from IPUMS
CPS and cover a balanced panel of couples where each individual's age is
between 25 and 65. The data cover the period between 1982 and 2021 and the
estimation is done over five year centered rolling windows.}
\end{figure}

The evolution of the estimates of the parameters that govern the dynamics
and the interdependence is shown in Figures \ref{FIGURE: Evolution of Panel
Gamma's over Time} and \ref{FIGURE: Evolution of Panel Rho's over Time}.
Specifically, we estimate the Schmidt-Strauss model in Table \ref{TABLE:
Including Lagged Working} for each ethnicity over rolling 5-year time-spans
and plotted the estimates of the $\gamma$'s and of $\rho$ against time.
Comparing the patterns in Figure \ref{FIGURE: Evolution of Panel Rho's over
Time} to the patterns in Figure \ref{FIGURE: Evolution of CS Rho's over Time}%
, we see that Black and Hispanic couples are more similar. This is
consistent with the finding in Table \ref{TABLE: Dynamic Cross Section By
Ethnicity}. Interestingly, the estimated $\rho$'s for Hispanics and for
Others are now much more stable over time, while the $\rho$ for Whites is
now trending up.

\begin{figure}[]
\caption{Evolution of $\protect\gamma$'s over Time by Household Ethnicity}
\label{FIGURE: Evolution of Panel Gamma's over Time}
\begin{center}
\includegraphics[width = 0.9\textwidth]{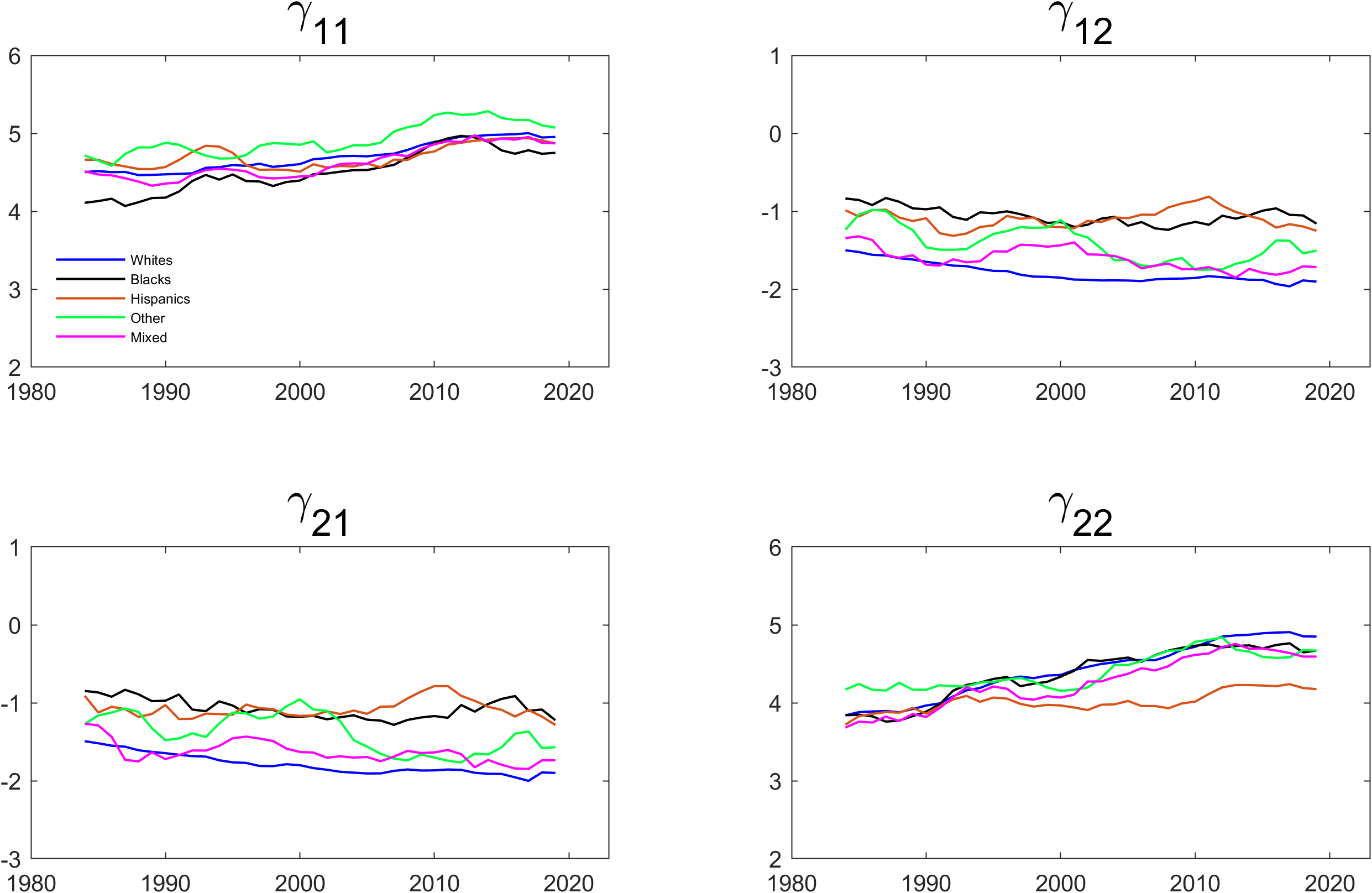}
\end{center}
\par
{\footnotesize \singlespacing  The dependent variable is working and the
parameters are estimated by maximum likelihood using the same specification
as in Table \ref{TABLE: Including Lagged Working}. The data are from IPUMS
CPS and cover a balanced panel of couples where each individual's age is
between 25 and 65. The data cover the period between 1982 and 2021 and the
estimation is done over five year centered rolling windows.}
\end{figure}

It is well-understood that it can be difficult to disentangle state
dependence (the causal dependence of a variable at one point in time from
its value in the previous period) from unobserved heterogeneity. Intuition
suggests that it is also difficult to distinguish between the effect of $%
\rho $ and the effect of unobserved heterogeneity that is correlated between
the husband and wife in the same household. These issues raise the question
of whether it is possible to semiparametrically identify $\rho$ and the
coefficients on the lagged dependent variables in a model that allows for
fixed effects. In the next section, we therefore investigate whether it is
possible to identify and estimate the parameters of a model that allows for
fixed effects in the dynamic Schmidt-Strauss framework. %

\section{Dynamic Panel Data Schmidt-Strauss Models with Fixed Effects}

\cite{HonoreKyriazidou2019a} adapt the Schmidt-Strauss model discussed in
Section \ref{SS Model} to a static panel data setting where each outcome can
also depend on unit-specific fixed effects. Specifically, %
they assume that%
\begin{align}
& P\left( \left. y_{1,it}=1\right\vert y_{2,it},\left\{
y_{1,is},y_{2,is}\right\} _{s<t},\left\{ x_{1,is}\right\} _{s=1}^{T},\left\{
x_{2,is}\right\} _{s=1}^{T},\alpha _{1,i},\alpha _{2,i}\right)  \notag \\
& =\Lambda \left( \alpha _{1,i}+x_{1,it}^{\prime }\beta _{1}+\rho
y_{2,it}\right)  \label{staticSS1}
\end{align}%
and%
\begin{align}
& P\left( \left. y_{2,it}=1\right\vert y_{1,it},\left\{
y_{1,is},y_{2,is}\right\} _{s<t},\left\{ x_{1,is}\right\} _{s=1}^{T},\left\{
x_{2,is}\right\} _{s=1}^{T},\alpha _{1,i},\alpha _{2,i}\right)
\label{staticSS2} \\
& =\Lambda \left( \alpha _{2,i}+x_{2,it}^{\prime }\beta _{2}+\rho
y_{1,it}\right)  \notag
\end{align}%
In this model, $\alpha _{1,i}$ and $\alpha _{2,i}$ are the fixed effects, $%
x_{1,it}$ and $x_{2,it}$ are strictly exogenous explanatory variables, and $%
\rho $ is the cross-equation dependence parameter, which as in \cite%
{SchmidtStrauss1975}, needs to be the same in the two equations given the
structure in equations (\ref{staticSS1}) and (\ref{staticSS2}).

Following \cite{SchmidtStrauss1975}, it can be shown that
\begin{align*}
& P\left( \left. y_{1,it}=c_{1},y_{2,it}=c_{2}\right\vert \left\{
y_{1,is},y_{2,is}\right\} _{s<t},\left\{ x_{1,is}\right\} _{s=1}^{T},\left\{
x_{2,is}\right\} _{s=1}^{T},\alpha _{1,i},\alpha _{2,i}\right) \\
& =\frac{\exp \left( c_{1}\left( \alpha _{1,i}+x_{1,it}^{\prime }\beta
_{1}\right) +c_{2}\left( \alpha _{2,i}+x_{2,it}^{\prime }\beta _{2}\right)
+c_{1}c_{2}\rho \right) }{1+\exp \left( \alpha _{1,i}+x_{1,it}^{\prime
}\beta _{1}\right) +\exp (\alpha _{2,i}+x_{2,it}^{\prime }\beta _{2})+\exp
\left( \alpha _{1,i}+x_{1,it}^{\prime }\beta _{1}+\alpha
_{2,i}+x_{2,it}^{\prime }\beta _{2}+\rho \right) }
\end{align*}%
for $c_{1},c_{2}\in \left\{ 0,1\right\} .$ \cite{HonoreKyriazidou2019a} show
that a conditional likelihood argument can be used to identify and estimate $%
\beta _{1}$, $\beta _{2}$, and $\rho $ with as few as $T=2$ time periods.
Indeed, $\rho $ can be allowed to be time dependent in equations (\ref%
{staticSS1}) and (\ref{staticSS2}).

\cite{HonoreKyriazidou2019a} also consider a vector autoregressive
simultaneous logit model:
\begin{eqnarray}
P\left( \left. y_{1,it}=1\right\vert
y_{2,it},y_{1,i}^{t-1},y_{2,i}^{t-1},\alpha _{1i},\alpha _{2i}\right)
&=&\Lambda \left( \alpha _{1i}+y_{1,it-1}\gamma _{11}+y_{2,it-1}\gamma
_{12}+\rho y_{2,it}\right) ,  \label{EQ: SS2} \\
P\left( \left. y_{2,it}=1\right\vert
y_{1,it},y_{1,i}^{t-1},y_{2,i}^{t-1},\alpha _{1i},\alpha _{2i}\right)
&=&\Lambda \left( \alpha _{2i}+y_{1,it-1}\gamma _{21}+y_{2,it-1}\gamma
_{22}+\rho y_{1,it}\right) .  \notag
\end{eqnarray}
This model is arguably the most relevant fixed effects specification for the
application in this paper. For each individual, we only use data from four
months, so with the exception of time-dummies, there is essentially no
exogenous variability in the explanatory variables over time. Moreover, we
use one time period to provide the initial conditions, and the effect of
time variables is probably not important over a three month period.\footnote{%
The 2008 financial crisis and the onset of the pandemic in 2020 are possible
exceptions to this.}

\cite{HonoreKyriazidou2019a} show that $\left( \gamma _{11},\gamma
_{12},\gamma _{21},\gamma _{22}\right) $ is identified in the model given in
equation (\ref{EQ: SS2}) with a total of four time periods (including the
one that delivers the initial condition). However, the conditioning argument
that leads to the identification eliminates the parameter $\rho $ along with
the fixed effects, $\alpha _{1i}$ and $\alpha _{2i}$. On the positive side,
this implies that one can allow the parameter $\rho $ in equation (\ref{EQ:
SS2}) to be individual-specific. On the other hand, $\rho $ may be the
parameter of interest in many applications,
including the one
considered here. This makes it problematic that the conditioning argument
eliminates it along with $\alpha _{1i}$ and $\alpha _{2i}$. In the next
subsection, we first generalize the results in \cite{HonoreKyriazidou2019a}
to show that using a conditional likelihood approach to eliminate $\alpha
_{1i}$ and $\alpha _{2i}$ in equation (\ref{EQ: SS2}) will also eliminate $%
\rho $ for all values of $T$. The conditional likelihood approach is then
illustrated empirically by obtaining estimates of the $\gamma$'s in equation
(\ref{EQ: SS2}) in the context of husbands' and wives' employment. Since the
simultaneity parameter, $\rho$, is not generally identified from a
conditional likelihood approach, we next consider a restricted version of
the model, in which the two individual fixed effects are the same, except
for an additive constant which is the same across all pairs. In our
application, we interpret this as a model with household specific fixed
effects. This model is also illustrated empirically.

\subsection{Conditional Likelihood for Dynamic Schmidt-Strauss Model with
Fixed Effects\label{SEC: General Case}}

The traditional approach to estimating nonlinear fixed effects models is to
find a sufficient statistic for the fixed effects, and then to construct a
conditional likelihood function conditioning on the sufficient statistic. By
construction, this conditional likelihood function will not depend on the
fixed effects and it may or may not depend on some or all of the parameters
of interest. In this subsection, we consider the conditional likelihood
approach for the model in equation (\ref{EQ: SS2}). This extends the
analysis in \cite{HonoreKyriazidou2019a}.

We consider a situation in which a pair of outcomes\footnote{%
In this and the following sections, we drop the subscript $i$ for simplicity.%
} $\left( y_{1,t},y_{2,t}\right) $ from equation (\ref{EQ: SS2}) are
observed for $T$ periods. We also assume that the initial condition, $\left(
y_{1,0},y_{2,0}\right) $, is observed. We denote the probability
distribution of $\left( y_{1,0},y_{2,0}\right) $ by $p\left(
y_{1,0},y_{2,0},\alpha _{1},\alpha _{2}\right) $, and we do not assume that
it is necessarily generated by the same model. For notational simplicity, we
let $z_{1,t}=\gamma _{11}y_{1,t}+\gamma _{12}y_{2,t}$ and $z_{2,t}=\gamma
_{21}y_{1,t}+\gamma _{22}y_{2,t}$.

With this, the probability of a particular sequence is
\begin{eqnarray*}
&&\frac{p\left( y_{1,0},y_{2,0},\alpha _{1},\alpha _{2}\right)
\prod\limits_{t=1}^{T}\exp \left( y_{1,t}\left( z_{1,t-1}+\alpha _{1}\right)
\right) \exp \left( y_{2,t}\left( z_{2,t}+\alpha _{2}\right) \right) \exp
\left( y_{1,t}y_{2,t}\rho \right) }{\prod\limits_{t=1}^{T}\left\{ 1+\exp
\left( z_{1,t-1}+\alpha _{1}\right) +\exp \left( z_{2,t-1}+\alpha
_{2}\right) +\exp \left( z_{1,t-1}+z_{2,t-1}+\alpha _{1}+\alpha _{2}+\rho
\right) \right\} } \\
&& \\
&=&\frac{p\left( y_{1,0},y_{2,0},\alpha _{1},\alpha _{2}\right) }{1+\exp
\left( z_{1,0}+\alpha _{1}\right) +\exp \left( z_{2,0}+\alpha _{2}\right)
+\exp \left( z_{1,0}+z_{2,0}+\alpha _{1}+\alpha _{2}+\rho \right) } \\
&&\frac{\prod\limits_{t=1}^{T}\exp \left( y_{1,t}\left( z_{1,t-1}+\alpha
_{1}\right) \right) \exp \left( y_{2,t}\left( z_{2,t}+\alpha _{2}\right)
\right) \exp \left( y_{1,t}y_{2,t}\rho \right) }{\prod\limits_{t=1}^{T-1}%
\left\{ 1+\exp \left( z_{1,t}+\alpha _{1}\right) +\exp \left( z_{2,t}+\alpha
_{2}\right) +\exp \left( z_{1,t}+z_{2,t}+\alpha _{1}+\alpha _{2}+\rho
\right) \right\} }.
\end{eqnarray*}

Now consider two different sequences of $\left\{ \left(
y_{1,t},y_{2,t}\right) \right\} _{t=1}^{T}$ with the same $\left(
y_{1,0},y_{2,0}\right) $. The probability of one of the sequences
conditional on observing one of the two depends on the ratio of the
probabilities for the two sequences. The key question is whether the
individual-specific effects cancel in that ratio.

In the numerator, the $\alpha $'s cancel if two sequences have the same $%
\sum_{t=1}^{T}y_{1,t}$ and the same $\sum_{t=1}^{T}y_{2,t}$. In the
denominator, each combination of $\left( y_{1,t}y_{2,t}\right) $ must appear
equally often. The latter is the same as saying that $%
\sum_{t=1}^{T-1}y_{1,t} $, $\sum_{t=1}^{T-1}y_{2,t}$, $%
\sum_{t=1}^{T-1}y_{1,t}y_{2,t}$ must be the same\footnote{%
On the other hand, it seems that the only way to generalize the conditioning
argument to a model that also allows for time varying explanatory variables
is to condition on equality of the explanatory variables across different
time periods. Without such a restriction, the fixed effects in the
denominators cannot cancel each other. \cite{ChountasKyriazidou2021} pursue
such a strategy for the conditional likelihood in a multinomial multivariate
model with discrete explanatory variables. In the case of continuous
explanatory variables, one may use the kernel weight approach introduced in
\cite{HonKyr00a}, although this would lead to an estimator that converges
slower than the usual $\sqrt{n}$.}. This suggests the sufficient statistic
\begin{equation*}
\left( y_{1,0},y_{2,0},\sum_{t=1}^{T-1}y_{1,t}\text{, }%
\sum_{t=1}^{T-1}y_{2,t},\sum_{t=1}^{T-1}y_{1,t}y_{2,t},y_{1,T},y_{2,T}\right)
\end{equation*}%
and the conditional likelihood function (for a given observation with fixed
effects $\alpha _{1}$ and $\alpha _{2}$) is therefore%
\begin{equation}
\mathcal{L=}\frac{\prod\limits_{t=1}^{T}\exp \left( y_{1,t}\left( \gamma
_{11}y_{1,t-1}+\gamma _{12}y_{2,t-1}\right) \right) \exp \left(
y_{2,t}\left( \gamma _{21}y_{1,t-1}+\gamma _{22}y_{2,t-1}\right) \right) \ }{%
\sum\limits_{\mathcal{B}}\prod\limits_{t=1}^{T}\exp \left( c_{t}\left(
\gamma _{11}c_{t-1}+\gamma _{12}d_{t-1}\right) \right) \exp \left(
d_{t}\left( \gamma _{21}c_{t-1}+\gamma _{22}d_{t-1}\right) \right) \ },
\label{EQ: Conditional Likelihood 1}
\end{equation}%
where $\mathcal{B}$ is the set of all sequences, $\{c_t,d_t\}_{t=0}^T$, such
that
\begin{equation*}
(c_{0},d_{0})=(y_{1,0},y_{2,0}),\sum_{t=1}^{T-1}c_{t}=%
\sum_{t=1}^{T-1}y_{1,t},\sum_{t=1}^{T-1}d_{t}=\sum_{t=1}^{T-1}y_{2,t}, \\
\sum_{t=1}^{T-1}c_{t}d_{t}=%
\sum_{t=1}^{T-1}y_{1,t}y_{2,t},(c_{T},d_{T})=(y_{1,T},y_{2,T}).
\end{equation*}
Note that not only does $\alpha $ drop out of the conditional likelihood,
but so does $\rho $. In other words, a conditional likelihood approach does
not identify $\rho $ for any $T$. Also note that the conditional likelihood
is constant if $T<3$, so at least three periods are needed in addition to
the one providing the initial conditions.

We finally note that the argument above is unchanged if one replaces $\gamma
_{11}$, $\gamma _{12}$, $\gamma _{21}$, $\gamma _{22}$, and $\rho$ with
functions of exogenous covariates as long as the functions do not change
over time. For example, in the application some of these parameters could be
functions of the level of education or of the presence of children.

\subsection{Empirical Illustration\label{SEC: Empirical Illustration.
General Case}}

In Table \ref{TABLE: Individual Fixed Effects Model By Ethnicity}, we
present the results from estimating $\gamma_{11}$, $\gamma _{12}$, $\gamma
_{21}$, and $\gamma _{22}$ using the conditional likelihood approach
discussed above for the full sample as well as by ethnicity. As one
might expect, these parameter are much lower in the fixed effects
specification than those reported in Table \ref{TABLE: Dynamic Cross Section
By Ethnicity},  where we do not allow for unobserved heterogeneity. Figure %
\ref{FIGURE: Fixed Effects Estimates of the Gammas Individual Heterogeneity }
shows the results of estimating the model on rolling 5-year sub-samples for
each ethnicity. The estimates are fairly stable over time, and not very
different across ethnicities. Overall, there is strong evidence that, after
controlling for fixed effects, an individual's own lagged employment has a
positive effect. The effect of the spouse's lagged employment tends to be
negative and smaller in magnitude. As a comparison, \cite%
{ChountasKyriazidou2021} estimate multinomial fixed effects model of
husbands and wives employment. They use quarterly data from the German
Socio-Economic panel for the years 2013-15 and four different labor states
(full time employment, part time employment, unemployment and out of labor
force), and find strong negative effects of the husband's lagged employment
on the wife, but mostly positive although statistically insignificant
effects of the wife's lagged employment on the husband.

\begin{table}[]
\caption{Estimates of Dynamic Schmidt-Strauss Model with Fixed Effects by
Household Ethnicity}
\label{TABLE: Individual Fixed Effects Model By Ethnicity}
\begin{center}
{\footnotesize
\begin{tabular*}{1.0\textwidth}{l@{\extracolsep{\fill}}rrrrrr}
\hline
& All & White & Black & Hispanic & Other & Mixed \\
$\gamma_{11}$ & $1.620 $\rlap{***} & $1.611 $\rlap{***} & $1.640 $\rlap{***}
& $1.601 $\rlap{***} & $1.684 $\rlap{***} & $1.761 $\rlap{***} \\
& ($0.014 $) & ($0.016 $) & ($0.061 $) & ($0.056 $) & ($0.080 $) & ($0.062 $)
\\
$\gamma_{12}$ & $-0.296 $\rlap{***} & $-0.338 $\rlap{***} & $-0.193 $%
\rlap{**} & $-0.078 $\rlap{} & $-0.030 $\rlap{} & $-0.268 $\rlap{***} \\
& ($0.022 $) & ($0.025 $) & ($0.085 $) & ($0.085 $) & ($0.113 $) & ($0.088 $)
\\
$\gamma_{21}$ & $-0.280 $\rlap{***} & $-0.311 $\rlap{***} & $-0.246 $%
\rlap{***} & $-0.039 $\rlap{} & $-0.079 $\rlap{} & $-0.302 $\rlap{***} \\
& ($0.021 $) & ($0.024 $) & ($0.086 $) & ($0.080 $) & ($0.112 $) & ($0.085 $)
\\
$\gamma_{22}$ & $1.357 $\rlap{***} & $1.350 $\rlap{***} & $1.415 $\rlap{***}
& $1.324 $\rlap{***} & $1.381 $\rlap{***} & $1.420 $\rlap{***} \\
& ($0.017 $) & ($0.019 $) & ($0.065 $) & ($0.058 $) & ($0.086 $) & ($0.067 $)
\\ \hline
\multicolumn{7}{c}{{*}** p$<$0.01, ** p$<$0.05, * p$<$0.1}%
\end{tabular*}
}
\end{center}
\par
{\footnotesize \singlespacing  The dependent variable is working and the
parameters are estimated maximizing the conditional likelihood in equation (%
\ref{EQ: Conditional Likelihood 1}). The data are from IPUMS CPS and cover a
balanced panel of couples where each individual's age is between 25 and 65.
The data cover the period between 1982 and 2021. }
\end{table}

\begin{figure}[]
\caption{Evolution of $\protect\gamma$'s over Time by Household Ethnicity
(Fixed Effects)}
\label{FIGURE: Fixed Effects Estimates of the Gammas Individual Heterogeneity
}
\begin{center}
\includegraphics[width = 0.9\textwidth]{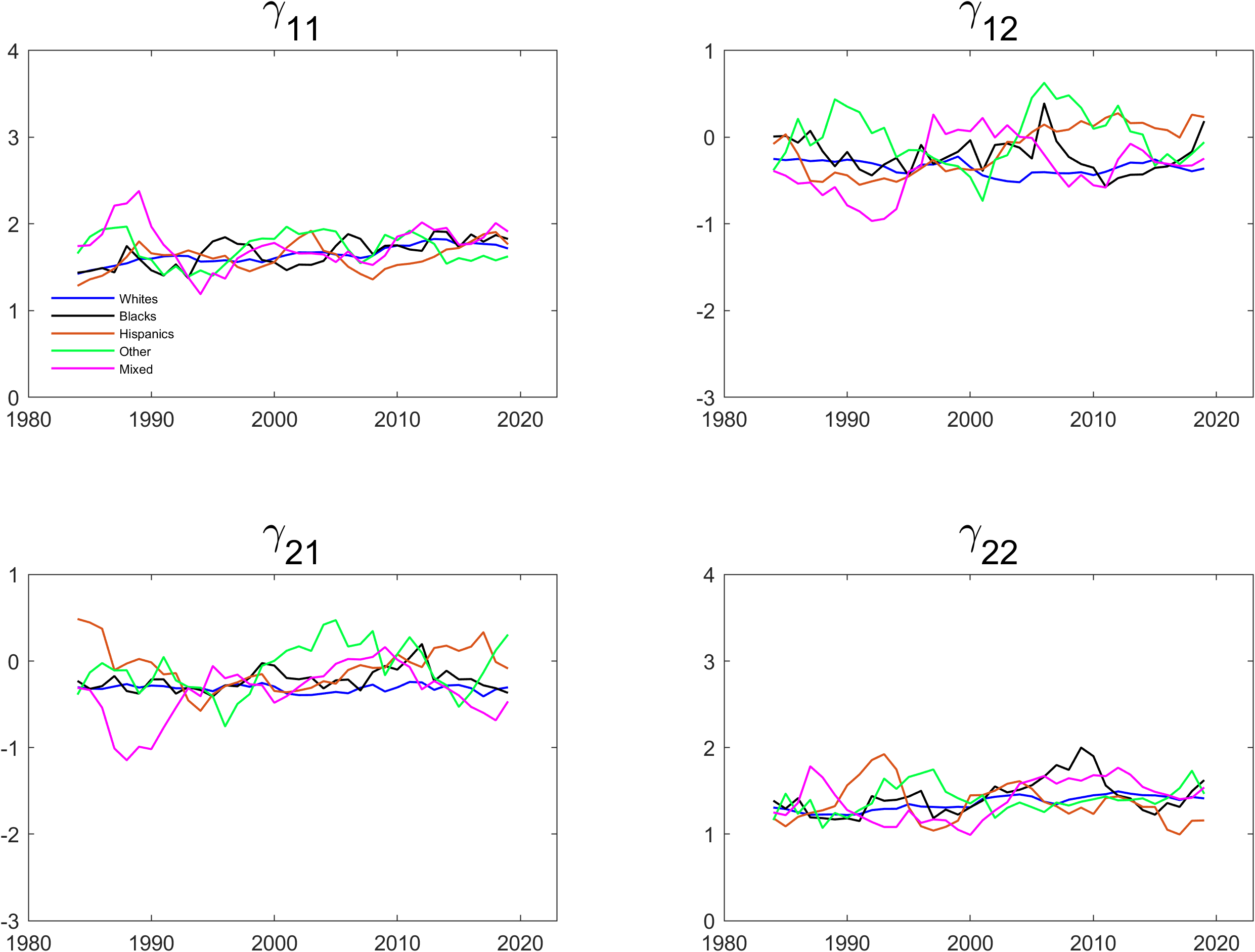}
\end{center}
\par
{\footnotesize \singlespacing The dependent variable is working and the
parameters are estimated maximizing the conditional likelihood in equation (%
\ref{EQ: Conditional Likelihood 1}). The data are from IPUMS CPS and cover a
balanced panel of couples where each individual's age is between 25 and 65.
The data cover the period between 1982 and 2021 and the estimation is done
over five year centered rolling windows.}
\end{figure}

\subsection{Conditional Likelihood for Dynamic Schmidt-Strauss Model with
Restricted Fixed Effects \label{SEC: Restricted Case}}

In this subsection, we investigate whether additional identification can be
obtained by assuming that $\alpha _{1}=\alpha $ and $\alpha _{2}=\alpha
+\kappa $ for some constant $\kappa $, which does not vary across units. Our
motivation is to see whether this will allow for identification of $\rho $.
In our application, we interpret this as a model with a family specific
fixed effect ($\alpha$) and a spouse specific level ($\kappa $).

As before, we consider a situation in which a pair of outcomes from equation
(\ref{EQ: SS2}) are observed for $T$ periods (in addition to period 0, which
delivers the initial condition). Again, we use the notation $z_{1,t}=\gamma
_{11}y_{1,t}+\gamma _{12}y_{2,t}$ and $z_{2,t}=\gamma _{21}y_{1,t}+\gamma
_{22}y_{2,t}$. With $\alpha _{2}=\alpha +\kappa $ , the probability of a
particular sequence becomes%
\begin{eqnarray*}
&&\frac{p\left( y_{1,0},y_{2,0},\alpha \right) \prod\limits_{t=1}^{T}\exp
\left( y_{1,t}\left( z_{1,t-1}+\alpha \right) \right) \exp \left(
y_{2,t}\left( z_{2,t}+\alpha +\kappa \right) \right) \exp \left(
y_{1,t}y_{2,t}\rho \right) }{\prod\limits_{t=1}^{T}\left\{ 1+\exp \left(
z_{1,t-1}+\alpha \right) +\exp \left( z_{2,t-1}+\alpha +\kappa \right) +\exp
\left( z_{1,t-1}+z_{2,t-1}+2\alpha +\kappa +\rho \right) \right\} } \\
&& \\
&=&\frac{p\left( y_{1,0},y_{2,0},\alpha \right) }{1+\exp \left(
z_{1,0}+\alpha \right) +\exp \left( z_{2,0}+\alpha +\kappa \right) +\exp
\left( z_{1,0}+z_{2,0}+2\alpha +\kappa +\rho \right) } \\
&&\frac{\prod\limits_{t=1}^{T}\exp \left( y_{1,t}\left( z_{1,t-1}+\alpha
\right) \right) \exp \left( y_{2,t}\left( z_{2,t}+\alpha +\kappa \right)
\right) \exp \left( y_{1,t}y_{2,t}\rho \right) }{\prod\limits_{t=1}^{T-1}%
\left\{ 1+\exp \left( z_{1,t}+\alpha \right) +\exp \left( z_{2,t}+\alpha
+\kappa \right) +\exp \left( z_{1,t}+z_{2,t}+2\alpha +\kappa +\rho \right)
\right\} }.
\end{eqnarray*}

As above, the key question is whether the unit-specific $a$'s cancel in the
ratio of the probabilities of two different sequences with the same initial
conditions. In the numerator, the $\alpha $'s cancel if the two sequences
have the same $\sum_{t=1}^{T}y_{1,t}+\sum_{t=1}^{T}y_{2,t}$. In the
denominator, each combination of $\left( y_{1,t},y_{2,t}\right) $ must
appear equally often\footnote{%
As was the case in Section \ref{SEC: General Case}, it seems that the only
way to generalize the conditioning argument to a model that also allows for
time varying variables is to condition on equality of the explanatory
variables across different time periods.}. The latter is the same as saying
that $\sum_{t=1}^{T-1}y_{1,t}$, $\sum_{t=1}^{T-1}y_{2,t}$, $%
\sum_{t=1}^{T-1}y_{1,t}y_{2,t}$ must be the same. This suggests the
sufficient statistic%
\begin{equation*}
\left(
y_{1,0},y_{2,0},\sum_{t=1}^{T-1}y_{1,t},\sum_{t=1}^{T-1}y_{2,t},%
\sum_{t=1}^{T-1}y_{1,t}y_{2,t},y_{1,T}+y_{2,T}\right)
\end{equation*}%
The difference from the case where the $\alpha $'s are unrestricted is that
we do not need to condition on $y_{1,T}$ and $y_{2,T}$, but only on the sum.
The implication is that a conditional likelihood approach will lead to more
sequences being compared to each other.

The conditional likelihood function (for a given individual) is
\begin{equation}
\mathcal{L}=\frac{\prod\limits_{t=1}^{T}\exp \left( y_{1,t}\left( \gamma
_{11}y_{1,t-1}+\gamma _{12}y_{2,t-1}\right) \right) \exp \left(
y_{2,t}\left( \gamma _{21}y_{1,t-1}+\gamma _{22}y_{2,t-1}+\kappa \right)
\right) }{\sum\limits_{\mathcal{B}}\prod\limits_{t=1}^{T}\exp \left(
c_{t}\left( \gamma _{11}c_{t-1}+\gamma _{12}d_{t-1}\right) \right) \exp
\left( d_{t}\left( \gamma _{21}c_{t-1}+\gamma _{22}d_{t-1}+\kappa \right)
\right) }  \label{EQ: Conditional Likelihood 2}
\end{equation}%
where $\mathcal{B}$ is the set of all sequences, $\{c_{t},d_{t}\}_{t=0}^{T}$%
, such that
\begin{equation*}
(c_{0},d_{0})=(y_{1,0},y_{2,0}),\sum_{t=1}^{T-1}c_{t}=%
\sum_{t=1}^{T-1}y_{1,t},\sum_{t=1}^{T-1}d_{t}=\sum_{t=1}^{T-1}y_{2,t},%
\sum_{t=1}^{T-1}c_{t}d_{t}=%
\sum_{t=1}^{T-1}y_{1,t}y_{2,t},c_{T}+d_{T}=y_{1,T}+y_{2,T}.
\end{equation*}%
Note that while $\alpha $ and $\rho $ drop out of this expression, $\kappa $
does not. Also note that this argument is unchanged if one replaces $\kappa $
with some function of predetermined covariates as long as the function does
not change over time. The same is true for the parameters $\gamma _{11}$, $%
\gamma _{12}$, $\gamma _{21}$, and $\gamma _{22}$.

\subsection{Empirical Illustration\label{SEC: Empirical Illustration.
Restricted Case}}

In Table \ref{TABLE: Household Fixed Effects Model By Ethnicity}, we present
the results from estimating $\gamma _{11}$, $\gamma _{12}$, $\gamma _{21}$,
and $\gamma _{22}$ using the conditional likelihood approach discussed above %
for the full sample as well as by ethnicity. The fixed effects
estimates are again lower than those reported in Table \ref{TABLE: Dynamic
Cross Section By Ethnicity}, which did not allow for unobserved
heterogeneity, but they are larger than the ones that were obtained when we
did not restrict the fixed effects for the husbands and the wives reported
in Table \ref{TABLE: Individual Fixed Effects Model By Ethnicity}. Since the
conditional likelihood in equation (\ref{EQ: Conditional Likelihood 2}) uses
more observations that the one in equation (\ref{EQ: Conditional Likelihood
1}), we would expect the estimated standard error to be smaller in Table \ref%
{TABLE: Household Fixed Effects Model By Ethnicity} than in Table \ref%
{TABLE: Individual Fixed Effects Model By Ethnicity}.

Figure \ref{FIGURE: Fixed Effects Estimates of the Gammas Household
Heterogeneity} shows the results of estimating the model on rolling 5-year
sub-samples for each ethnicity. The estimates are fairly stable over time,
and not very different across ethnicities.

\begin{table}[]
\caption{Estimates of Dynamic Schmidt-Strauss Model with Restricted Fixed
Effects by Household Ethnicity}
\label{TABLE: Household Fixed Effects Model By Ethnicity}
\par
\begin{center}
{\footnotesize
\begin{tabular*}{1.0\textwidth}{l@{\extracolsep{\fill}}rrrrrr}
\hline
& All & Whites & Blacks & Hispanics & Other & Mixed \\
$\gamma_{11}$ & $2.385 $\rlap{***} & $2.374 $\rlap{***} & $2.364 $\rlap{***}
& $2.403 $\rlap{***} & $2.436 $\rlap{***} & $2.485 $\rlap{***} \\
& ($0.014 $) & ($0.016 $) & ($0.058 $) & ($0.059 $) & ($0.078 $) & ($0.060 $)
\\
$\gamma_{12}$ & $-1.511 $\rlap{***} & $-1.542 $\rlap{***} & $-1.409 $%
\rlap{***} & $-1.318 $\rlap{***} & $-1.298 $\rlap{***} & $-1.505 $\rlap{***}
\\
& ($0.015 $) & ($0.018 $) & ($0.063 $) & ($0.060 $) & ($0.081 $) & ($0.065 $)
\\
$\gamma_{21}$ & $-1.538 $\rlap{***} & $-1.576 $\rlap{***} & $-1.392 $%
\rlap{***} & $-1.403 $\rlap{***} & $-1.310 $\rlap{***} & $-1.485 $\rlap{***}
\\
& ($0.015 $) & ($0.016 $) & ($0.060 $) & ($0.060 $) & ($0.078 $) & ($0.061 $)
\\
$\gamma_{22}$ & $2.263 $\rlap{***} & $2.285 $\rlap{***} & $2.231 $\rlap{***}
& $2.097 $\rlap{***} & $2.206 $\rlap{***} & $2.315 $\rlap{***} \\
& ($0.016 $) & ($0.018 $) & ($0.064 $) & ($0.059 $) & ($0.083 $) & ($0.065 $)
\\ \hline
\multicolumn{7}{c}{{*}** p$<$0.01, ** p$<$0.05, * p$<$0.1}%
\end{tabular*}
}
\end{center}
\par
{\footnotesize \singlespacing The dependent variable is working and the
parameters are estimated maximizing the conditional likelihood in equation (%
\ref{EQ: Conditional Likelihood 2}). The data are from IPUMS CPS and cover a
balanced panel of couples where each individual's age is between 25 and 65.
The data cover the period between 1982 and 2021. }
\end{table}

\begin{figure}[h]
\caption{Evolution of $\protect\gamma$'s over Time by Household Ethnicity
(Restricted Fixed Effects)}
\label{FIGURE: Fixed Effects Estimates of the Gammas Household Heterogeneity}
\begin{center}
\includegraphics[width = 0.9\textwidth]{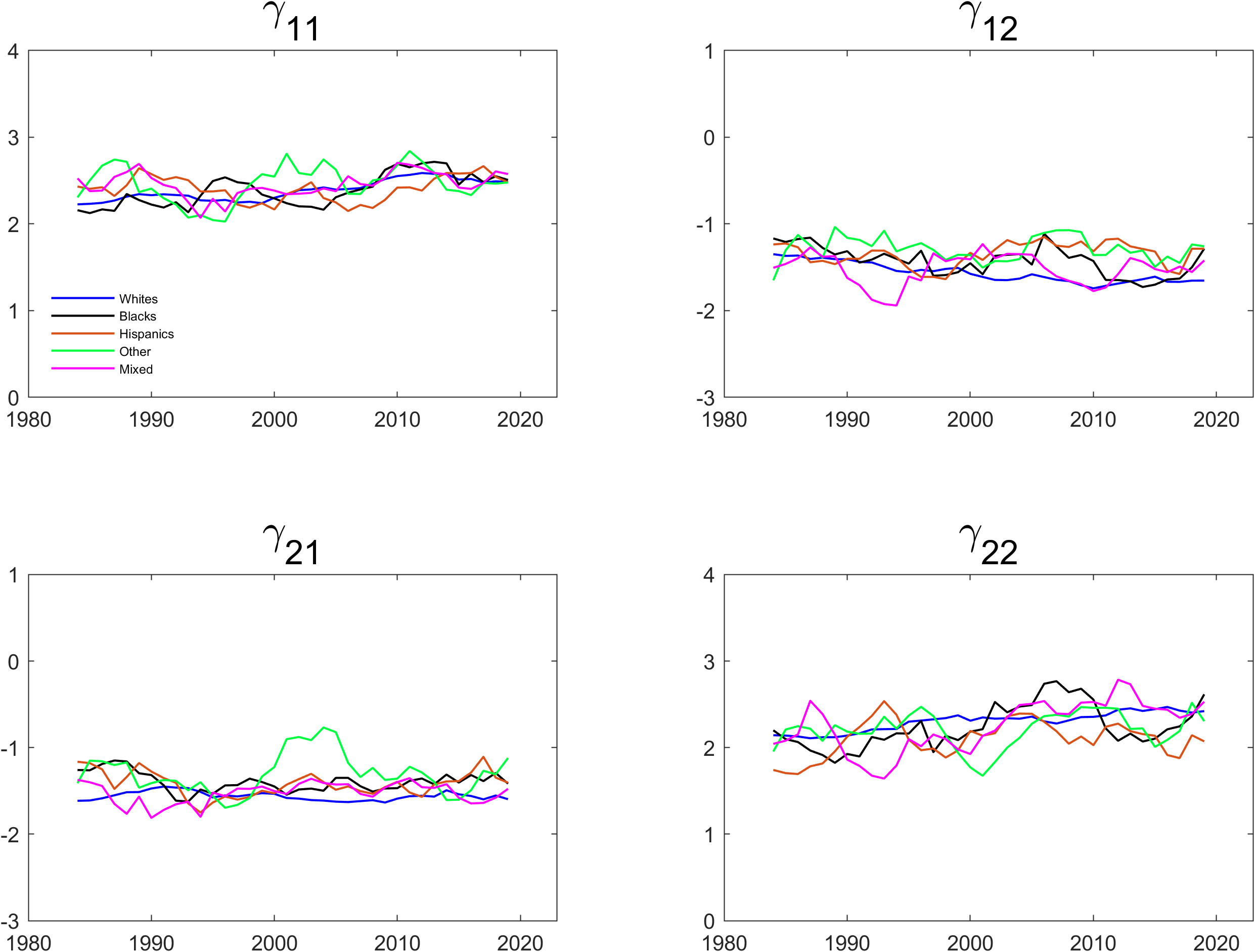}
\end{center}
\par
{\footnotesize \singlespacing The dependent variable is working and the
parameters are estimated maximizing the conditional likelihood in equation (%
\ref{EQ: Conditional Likelihood 2}). The data are from IPUMS CPS and cover a
balanced panel of couples where each individual's age is between 25 and 65.
The data cover the period between 1982 and 2021 and the estimation is done
over five year centered rolling windows.}
\end{figure}

\section{Moment Conditions for the Dynamic Schmidt-Strauss Model with Fixed
Effects}

In panel data models with fixed effects, it is sometimes possible to
construct moment conditions that do not depend on the fixed effects. When
that is the case, one can consider estimating the common parameters of the
model by generalized method of moments. The dynamic linear panel data model
is a simple example of this; see, for example \cite{AndHsi81} or \cite%
{Holtz-EakinNeweyRosen88}. Applications of this idea to nonlinear models
include \cite{Honore92}, \cite{Kyriazidou01}, \cite{Hu2002} and \cite%
{kitazawa2013exploration}.\footnote{%
The maximum score estimator in \cite{Manski87} can be motivated in terms of
moment inequalities.} \cite{Bonhomme2012} proposes a general approach for
constructing such moment conditions and \cite{HonoreWeidner2020} develop a
specific numeric strategy for determining whether such moment conditions can
be constructed in particular models with discrete outcomes. In this section,
we report the results from applying the approach in \cite{HonoreWeidner2020}
to determine whether there are moments that can be used to
identify
and estimate $\rho$ in a Schmidt-Strauss model with lagged dependent
variables and fixed effects.

We consider two versions of the model
\begin{eqnarray*}
&&P\left( \left. y_{1,t}=c_{1},y_{2,t}=c_{2}\right\vert \left\{
y_{1,s},y_{2,s}\right\} _{s<t},\left\{ x_{1,s}\right\} _{s=1}^{T},\left\{
x_{2,s}\right\} _{s=1}^{T},\alpha _{1},\alpha _{2}\right) \\
&=&\frac{\exp \left( c_{1}\left( z_{1,t}+\alpha _{1}\right) +c_{2}\left(
z_{2,t}+\alpha _{2}\right) +c_{1}c_{2}\rho \right) }{1+\exp \left(
z_{1,t}+\alpha _{1}\right) +\exp (z_{2,t}+\alpha _{2})+\exp \left(
z_{1,t}+\alpha _{1}+z_{2,t}+\alpha _{2}+\rho \right) }
\end{eqnarray*}%
for $t=1,2,3$ and $c_{1},c_{2}\in \left\{ 0,1\right\} \ $, where $%
z_{1,t}=x_{1,t}^{\prime }\beta _{1}+y_{1,t-1}\gamma _{11}+y_{2,t-1}\gamma
_{12}$ and $z_{2,t}=x_{2,t}^{\prime }\beta _{2}+y_{1,t-1}\gamma
_{21}+y_{2,t-1}\gamma _{22}.$ In one version, $\alpha _{1}$ and $\alpha _{2}$
are unrestricted  as in Section \ref{SEC: General Case}, while the
other version restricts them to be identical except for an additive constant
as in Section \ref{SEC:
Restricted Case}.
Note that these are the same models as in Sections \ref{SEC: General Case} and \ref{SEC:
Restricted Case}, except that we here  allow for
strictly exogenous covariates.

Table \ref{TABLE: Counting Moment Conditions} reports the number of moment
conditions for each of the two versions of the model when one has 3, 4 or 5
time periods of observations in addition to the one that provides the
initial conditions. The data used in this paper has a total of four
consecutive time periods, and the results for $T=3$ are therefore the
relevant ones here. In the empirical illustration in Sections \ref%
{SEC: Empirical Illustration. General Case} and \ref{SEC: Empirical
Illustration. Restricted Case}, we have no strictly exogenous time-varying
explanatory variables, so according to the calculation reported in Table \ref%
{TABLE: Counting Moment Conditions}, there will be no moment conditions that
depend on $\rho$ when the fixed effects are left unrestricted. On the other
hand, there will be six moment conditions for each initial condition when
the fixed effects are restricted. With more than three time periods (in
addition to the one providing the initial conditions), the results suggest
that there are moment conditions that depend on $\rho$ even when the fixed
effects are unrestricted. While introducing explanatory variables changes
the number of moment conditions, it does not change the answer to the
question of whether there exist moment conditions that depend on $\rho$ for
a given value of $T$.

\begin{table}[]
\caption{The Number of Moment Conditions in the Dynamic Schmidt-Strauss
Model with Fixed Effects}
\label{TABLE: Counting Moment Conditions}
\begin{center}
\begin{tabular}{@{}c||@{\;}c||c||c}
& $T=3$ & $T=4$ & $T=5$ \\ \hline
$x_{k,t}=0$, unrestricted $(\alpha_{1},\alpha_{2})$ & 24 / 21 / 0 & 180 /
136 / 4 & 900 / 534 / 16 \\[5pt]
$x_{k,t}=0$, restricted $\alpha_{2} = \alpha_{1} + \kappa$ & 45 / 42 / 6 &
229 / 185 / 18 & 989 / 623 / 36 \\[5pt]
$x_{k,t} \neq 0$, unrestricted $(\alpha_{1},\alpha_{2})$ & 4 / 4 / 0 & 120 /
120 / 64 & 780 / 780 / 256 \\[5pt]
$x_{k,t} \neq 0$, restricted $\alpha_{2} = \alpha_{1} + \kappa$ & 45 / 45 /
16 & 229 / 229 / 48 & 989 / 989 / 96%
\end{tabular}%
\end{center}
\par
{\footnotesize \singlespacing Results from the numerical counting of moment
conditions for the dynamic simultaneous logit are reported. Four different
model specifications are considered: additional exogenous regressors are
present ($x_{k,t} \neq 0 $) or not ($x_{k,t} = 0$), and the fixed effects $%
(\alpha_{1},\alpha_{2})$ are unrestricted or restricted ($\alpha_{2} =
\alpha_{1} + \kappa$). For each of those four specifications and each value
of $T$ we report $n_{\mathrm{tot}} \, / \, n_{\mathrm{para}} \, / \,
n_{\rho} $, where $n_{\mathrm{tot}}$ is the total number of moment
conditions available, $n_{\mathrm{para}}$ is the number of moment conditions
available that depend on any of the common parameters ($\gamma_{11}$, $%
\gamma_{12}$, $\gamma_{21}$, $\gamma_{22}$ $\beta_1$, $\beta_2$, $\rho$, $%
\kappa$), and $n_{\rho}$ is the number of moment conditions available that
depend on the parameter $\rho$. All results are for one fixed value of the
initial condition $(y_{1,0},y_{2,0})$, but the number of moment conditions
is independent from the initial condition. Notice that for $T=3$ and
unrestricted $(\alpha_{1},\alpha_{2})$ we have $n_{\rho} = 0$, and in
general we believe that the parameter is not identified in that case.
However, for either $T>3$ or restricted $\alpha_{2} = \alpha_{1} + \kappa $
we find that $n_{\rho} > 0$ and the parameter $\rho$ can be identified and
estimated from those moment conditions. }
\end{table}

\subsection{Moment Conditions For $\protect\rho$ \label{Some Moment
Conditions}}

It is not always easy to derive analytical expressions for the moment
conditions. For the empirical application in Sections \ref{SEC: Empirical
Illustration. General Case} and \ref{SEC: Empirical Illustration. Restricted
Case} of this paper, $T$ is three and there are no strictly exogenous
time-varying explanatory variables. In order to make statements about $\rho $%
, we therefore have to limit attention to the model in which the fixed
effect is household specific in the sense that $\alpha _{2}=\alpha
_{1}+\kappa $.

As mentioned above, there will be a total of 45 moment conditions in this
case. One can write these as six that depend on $\rho$, 36 that depend on
some of the common parameters in the model, but not on $\rho$, and three
that do not depend on any of the parameters in the model. In principle, one
may need to use all of these moments to construct an efficient GMM
estimator. On the other hand, we can already identify the $\gamma$'s and $%
\kappa$ from the conditional likelihood approach in Section \ref{SEC:
Restricted Case}, so we only need to use one moment\footnote{%
Subject to an identification condition that guarantees that the moment
condition has a unique solution for $\rho$.} that depends on $\rho$ in order
to (inefficiently) estimate $\rho$. We therefore focus on finding the six
linearly independent moment conditions that depend on $\rho$. Unfortunately,
these will not be unique. For example, adding a linear combination of moment
conditions that do not depend on $\rho $ to one of the six that do, will
leave us with six linearly independent moment conditions that depend on $%
\rho $. This also means that some of the moment conditions can be extremely
complicated.

Fortunately, it turns out that for the model considered here, one can find
six linearly independent moment conditions (for each initial condition)
which all depend on $\rho $, and where each only depends on five of the 64
possible sequences. They are given in the Appendix, and we use those to
estimate $\rho $ in the next subsection. These moment conditions are linear
in $\exp \left( \rho \right) $.

\subsection{Empirical Illustration}

In this subsection, we illustrate how the method of moments approach
discussed above can be used to estimate $\rho $  in the dynamic
Schmidt-Strauss model with restricted fixed effects. We proceed in two
steps. We first estimate the $\gamma $'s and $\kappa $ using the conditional
likelihood approach. We then fix the $\gamma $'s and $\kappa $ at those
estimates and estimate $\rho $ by generalized method of moments using the
moment conditions in the Appendix. As weighting matrix, we use the inverse
of a diagonal matrix that has the variance of the moments evaluated at $\rho
=0$ in the diagonal. This choice is arbitrary and may lead to statistical
inefficiency, but $\rho =0$ is a natural benchmark, and the hope is that
using a diagonal matrix will alleviate small sample issues resulting from
estimation of an efficient weighting matrix.\footnote{%
While the overall sample is large, each of the moment only depends on
specific sequences that comprise very small fraction of the observations. In
our application, these fractions ranged from less than 0.1\% to 3\%.} Since
the moment conditions are linear in $\exp \left( \rho \right) $, the GMM\
objective function will be quadratic in $\exp \left( \rho \right) $. This
implies that it is numerically well behaved and that $\rho $ is actually
identified from it. On the other hand, the solution for $\exp \left( \rho
\right) $, can sometimes be negative in finite samples. For the estimation
below, we search over values of $\rho $ between $-2$ and $4$.

The results of the estimation of $\rho$ are presented in Table \ref{TABLE:
Rho Household Fixed Effects Model By Ethnicity}. Compared to the estimates
of $\rho$ presented in Table \ref{TABLE: Dynamic Cross Section By Ethnicity}%
, the fixed effects estimates are much smaller. This suggests that the
household specific fixed effect captures much more of the intra-household
correlation than the observed characteristics.

\begin{table}[]
\caption{GMM Estimation of $\protect\rho$ by Household Ethnicity (Restricted
Fixed Effects)}
\label{TABLE: Rho Household Fixed Effects Model By Ethnicity}
\par
\begin{center}
{\footnotesize
\begin{tabular*}{1.0\textwidth}{l@{\extracolsep{\fill}}rrrrrr}
\hline
& All & Whites & Blacks & Hispanics & Other & Mixed \\
$\rho$ & $1.260 $\rlap{***} & $1.420 $\rlap{***} & $0.360 $\rlap{*} & $0.550
$\rlap{***} & $0.730 $\rlap{***} & $0.960 $\rlap{***} \\
& ($0.041$) & ($0.052$) & ($0.211$) & ($0.156$) & ($0.207$) & ($0.163$) \\
\hline
\multicolumn{7}{c}{{*}** p$<$0.01, ** p$<$0.05, * p$<$0.1}%
\end{tabular*}
}
\end{center}
\par
{\footnotesize \singlespacing The dependent variable is working.
The parameter $\rho$ is estimated by generalized method of moments using the moment
conditions in the Appendix, and the $\gamma$'s and $\kappa$ by the conditional likelihood method in Section \ref{SEC: Restricted Case}%
. The data are from IPUMS CPS and cover a balanced panel of couples where
each individual's age is between 25 and 65. The data cover the period
between 1982 and 2021. Standard errors are calculated via the bootstrap.
Bootstrap estimates of the vector of $\gamma$'s are obtained by
bootstrapping their influence function. Bootstrap estimates of $\rho$ are
then calculated using GMM after recalculating the weighting matrix.}
\end{table}

\begin{figure}[]
\caption{Evolution of $\protect\rho$ over Time by Household Ethnicity
(Restricted Fixed Effects)}
\label{FIGURE: Evolution of FE Rho's over Time}
\begin{center}
\includegraphics[width = 0.45\textwidth]{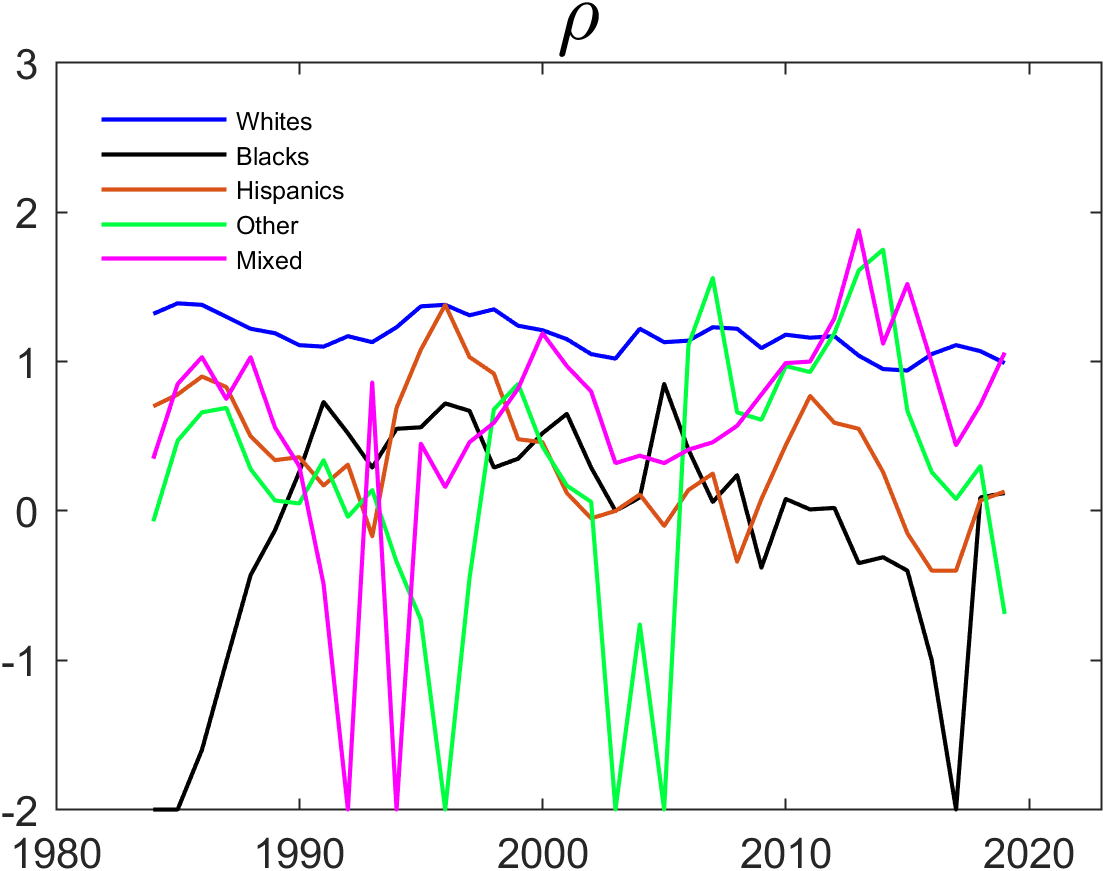}
\end{center}
\par
{\footnotesize \singlespacing The dependent variable is working.
$\rho$ is estimated by generalized method of moments using the moment
conditions in the appendix, and the remaining parameters by the conditional likelihood method in Section \ref{SEC: Restricted Case}%
. The data are from IPUMS CPS and cover a balanced panel of couples where
each individual's age is between 25 and 65. The data cover the period
between 1982 and 2021 and the estimation is done over five year centered
rolling windows. }
\end{figure}

Figure \ref{FIGURE: Evolution of FE Rho's over Time} presents the results of
estimating $\rho$ separately for each ethnicity over rolling 5-year periods.
The estimates for Whites seem fairly stable over time and are statistically
significantly different from 0 in all time periods.\footnote{%
The p-value for the test is less than 1\% in all cases. All test referred to
in this paragraph are based on estimating $\exp(\rho)$ without imposing that
it is positive, and then testing whether it differs from $\exp(0)$. The
reason is that when we estimate $\rho$, we sometimes obtain a point estimate
at the lower bound of the parameter space.} When testing at a 5\% level of
significance, the estimates for the other ethnicities are statistically
significantly different from 0 in only six of 144 cases (four for Blacks and
two for Others).

\section{Dynamic Schmidt-Strauss Models with Correlated Random Effects}

The calculations reported above establish that $\left( \gamma _{11},\gamma
_{12},\gamma _{21},\gamma _{22},\kappa ,\rho \right) $ in the model in
Section \ref{SEC: Restricted Case}\ is semiparametrically identified without
assumptions on $\alpha $. In such cases, \cite{Wooldridge2005} has proposed
estimating $\left( \gamma _{11},\gamma _{12},\gamma _{21},\gamma
_{22},\kappa ,\rho \right) $ by maximum likelihood conditional on the
initial observations, $(y_{1,0},y_{2,0})$, after modeling the distribution
of $\alpha $ conditional on those initial observations.
This approach is
in the spirit of \cite{Mundlak1978} and \cite{Chamberlain82:JOE} and is
known as a correlated random effects approach. See also \cite%
{Wooldridge2019}. If the conditional distribution of $\alpha $ %
given the initial conditions is sufficiently flexible, then one
might interpret this approach as a semiparametric sieve maximum likelihood
estimator.

Table \ref{TABLE: CRE Model By Ethnicity} shows the estimates of $\left(
\gamma _{11},\gamma _{12},\gamma _{21},\gamma _{22},\kappa ,\rho \right) $
that we obtain from the correlated random effects approach after modelling $%
\alpha $ conditional on $(y_{1,0},y_{2,0})$ as%
\begin{equation}
\alpha =\delta _{0}+y_{1,0}\delta _{1}+y_{2,0}\delta
_{2}+y_{1,0}y_{2,0}\delta _{3}+\nu ,\qquad \nu \sim N\left( 0,\sigma
^{2}\right) .  \label{EQ: CRE1}
\end{equation}

\begin{table}[tbp]
\caption{Estimates of Dynamic Schmidt-Strauss Model with Correlated Random
Effects by Household Ethnicity}
\label{TABLE: CRE Model By Ethnicity}
\par
\begin{center}
{\footnotesize
\begin{tabular*}{1.0\textwidth}{l@{\extracolsep{\fill}}rrrrrr}
\hline
& All & Whites & Blacks & Hispanics & Other & Mixed \\
$\gamma_{11}$ & $3.401 $*** & $3.414 $*** & $3.115 $*** & $3.297 $*** & $%
3.608 $*** & $3.405 $*** \\
& $( 0.009) $ & $( 0.010) $ & $( 0.038) $ & $( 0.036) $ & $( 0.048) $ & $(
0.035) $ \\
$\gamma_{12}$ & $-2.686 $*** & $-2.752 $*** & $-2.274 $*** & $-2.188 $*** & $%
-2.534 $*** & $-2.629 $*** \\
& $( 0.010) $ & $( 0.011) $ & $( 0.045) $ & $( 0.042) $ & $( 0.057) $ & $(
0.041) $ \\
$\gamma_{21}$ & $-2.859 $*** & $-2.933 $*** & $-2.320 $*** & $-2.404 $*** & $%
-2.654 $*** & $-2.779 $*** \\
& $( 0.011) $ & $( 0.012) $ & $( 0.046) $ & $( 0.047) $ & $( 0.062) $ & $(
0.045) $ \\
$\gamma_{22}$ & $3.325 $*** & $3.383 $*** & $3.050 $*** & $2.898 $*** & $%
3.336 $*** & $3.278 $*** \\
& $( 0.010) $ & $( 0.011) $ & $( 0.039) $ & $( 0.036) $ & $( 0.050) $ & $(
0.038) $ \\
$\rho$ & $0.866 $*** & $1.023 $*** & $0.013 $ & $0.007 $ & $0.496 $*** & $%
0.703 $*** \\
& $( 0.011) $ & $( 0.012) $ & $( 0.050) $ & $( 0.048) $ & $( 0.066) $ & $(
0.045) $ \\
$\kappa$ & $0.855 $*** & $0.854 $*** & $0.354 $*** & $1.333 $*** & $0.894 $%
*** & $0.760 $*** \\
& $( 0.007) $ & $( 0.008) $ & $( 0.027) $ & $( 0.027) $ & $( 0.035) $ & $(
0.029) $ \\ \hline
\multicolumn{7}{c}{{*}** p$<$0.01, ** p$<$0.05, * p$<$0.1}%
\end{tabular*}
}
\end{center}
\par
{\footnotesize \singlespacing  The dependent variable is working and the
parameters are estimated by maximizing the likelihood function conditional
on the initial conditions and under the assumption that $\alpha$ is
distributed as in equation (\ref{EQ: CRE1}). The data are from IPUMS CPS and
cover a balanced panel of couples where each individual's age is between 25
and 65. The data cover the period between 1982 and 2021. }
\end{table}

The estimates of $\left( \gamma _{11},\gamma _{12},\gamma _{21},\gamma
_{22}\right) $ in Table \ref{TABLE: CRE Model By Ethnicity} are larger in
magnitude than those reported in Table \ref{TABLE: Household Fixed Effects
Model By Ethnicity}, but the overall pattern is similar. The coefficients on
one's own past employment for women and for men, $\gamma _{11\text{ }}$and $%
\gamma _{22}$, are positive and of the same magnitude, and the coefficients
on the spouse's past employment for women and for men, $\gamma _{12\text{ }}
$and $\gamma _{21}$, are negative and of the same magnitude. Moreover, these
coefficients are estimated to be fairly similar across ethnicities. The
estimates for $\rho $ in Table \ref{TABLE: CRE Model By Ethnicity} show the
same pattern as the estimates in Table \ref{TABLE: Rho Household Fixed
Effects Model By Ethnicity}. Whites have the largest coefficient, while the
estimates for Blacks and Hispanics are much lower. The parameters estimated
based on the correlated random effects approach have less sampling
uncertainty than the fixed effects estimators in Section \ref{SEC:
Restricted Case} (presumably because they are based on additional
assumptions).

\begin{figure}[h]
\caption{Evolution of $\protect\gamma$'s over Time by Household Ethnicity
(Correlated Random Effects)}
\label{FIGURE: CRE Estimates of the Gammas Household Heterogeneity}
\begin{center}
\includegraphics[width = 0.9\textwidth]{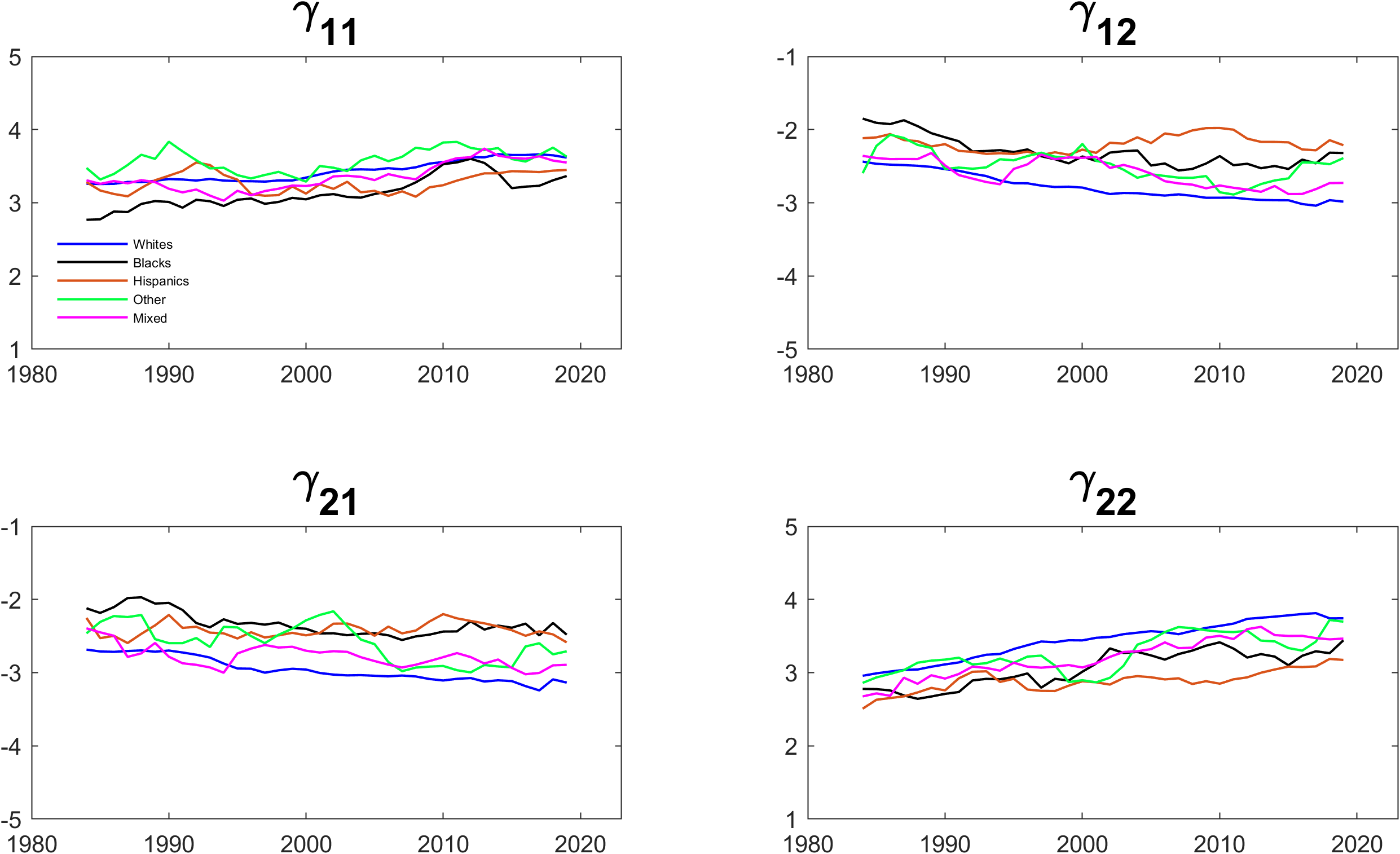}
\end{center}
\par
{\footnotesize \singlespacing The dependent variable is working and the
parameters are estimated by maximizing the conditional likelihood in
equation (\ref{EQ: Conditional Likelihood 2}). The data are from IPUMS CPS
and cover a balanced panel of couples where each individual's age is between
25 and 65. The data cover the period between 1982 and 2021 and the
estimation is done over five year centered rolling windows.}
\end{figure}

\begin{figure}[tbp]
\caption{Evolution of $\protect\rho$ over Time by Household Ethnicity
(Correlated Random Effects)}
\label{FIGURE: Evolution of CFE Rho's over Time}
\begin{center}
\includegraphics[width = 0.45\textwidth]{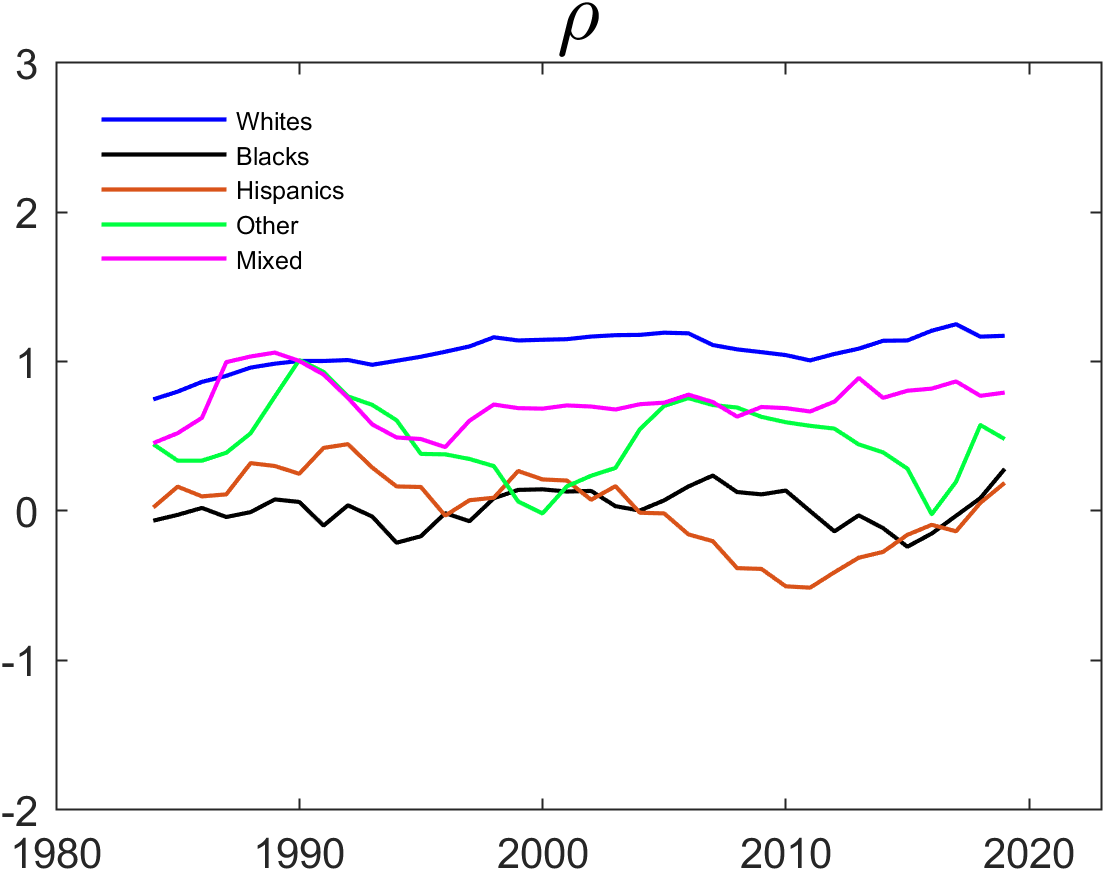}
\end{center}
\par
{\footnotesize \singlespacing The dependent variable is working and the
parameters are estimated by the correlated random effects approach.
The data are from IPUMS CPS and cover a balanced panel of couples where each
individual's age is between 25 and 65. The data cover the period between
1982 and 2021 and the estimation is done over five year centered rolling
windows. }
\end{figure}

Figures \ref{FIGURE: CRE Estimates of the Gammas Household Heterogeneity}
and \ref{FIGURE: Evolution of CFE Rho's over Time} show the results of
estimating the model using rolling 5-year sub-samples for each ethnicity.
The estimates are fairly stable over time, and not very different across
ethnicities. In terms of patterns, the results from estimating the $\gamma$%
's presented in Figure \ref{FIGURE: CRE Estimates of the Gammas Household
Heterogeneity} mainly differ from the fixed effects estimates presented in
Figure \ref{FIGURE: Fixed Effects Estimates of the Gammas Household
Heterogeneity} by displaying a clearer upward trend in the husband's
coefficient on his own past employment, $\gamma_{22}$. The estimates also
tend to have less sampling uncertainty. Again, this is to be expected
because the correlated random effects approach imposes additional structure
relative to the fixed effects approach. The correlated random effects
estimates of the $\rho$'s presented in Figure \ref{FIGURE: Evolution of CFE
Rho's over Time} are also noticeably less volatile than the GMM estimates in
Figure \ref{FIGURE: Evolution of FE Rho's over Time}.

The fact that the correlated random effects approach is associated with less
sampling uncertainty than the conditional likelihood approach comes at a
price: If the parametric form for the individual specific effect is
misspecified then the estimator can be inconsistent. For a given simple data
generating process, one can gauge the importance of this by calculating the
maximizer of the limiting (the expected) log-likelihood function for the
conditional random effects model. This is especially easy if the data
generating process for the fixed effects is discrete because the limiting
objective function becomes a sum rather than an integral in that case. This
maximizer of the limiting log-likelihood function will be the probability
limit of the conditional random effects estimator. To illustrate this, let $%
\left( \gamma _{11},\gamma _{12},\gamma _{21},\gamma _{22},\rho ,\kappa
\right) =\left( 2.5,-1.5,-1.5,2.5,1,2\right) $ and assume that $y_{1,0}$ and
$y_{2,0} $ are independent and equal to 1 with probability $\frac{1}{2}$. We
can then maximize the limiting objective functions for the correlated random
effects that assumes (\ref{EQ: CRE1}) under the following distributions for $%
\alpha $:

\begin{itemize}
\item Correctly specified: $\alpha =-1+y_{1,0}+y_{2,0}+\nu $, where $\nu
\sim N\left( 0,1\right) .$

\item {Discrete, but approximately normal: $\alpha =\eta $ where }

\begin{itemize}
\item $P\left( \left. \eta =-d\right\vert y_{1,0},y_{2,0}\right) =P\left(
\left. \eta =d\right\vert y_{1,0},y_{2,0}\right) =\Phi \left( -1.5\right) ,$

\item $P\left( \left. \eta =-1\right\vert y_{1,0},y_{2,0}\right) =P\left(
\left. \eta =1\right\vert y_{1,0},y_{2,0}\right) =\Phi \left( 1.5\right)
-\Phi \left( 0.5\right) $, and

\item $P\left( \left. \eta =0\right\vert y_{1,0},y_{2,0}\right) =\Phi \left(
0.5\right) -\Phi \left( -0.5\right) $,
\end{itemize}

where $\Phi $ is the standard normal cumulative distribution function and $%
d\approx 1.9662$ is chosen such that $\eta $ has variance 1.

\item Discrete, asymmetric: $P\left( \left. \alpha =3\right\vert
y_{1,0},y_{2,0}\right) =\frac{1}{4}$, $P\left( \left. \alpha =-1\right\vert
y_{1,0},y_{2,0}\right) =\frac{3}{4}$.

\item Heteroskedastic: $P\left( \left. \alpha =-\sqrt{2+2y_{1,0}}\right\vert
y_{1,0},y_{2,0}\right) =P\left( \left. \alpha =\sqrt{2+2y_{1,0}}\right\vert
y_{1,0},y_{2,0}\right) =\frac{1}{2}$.

\item Very heteroskedastic:$P\left( \left. \alpha =-\sqrt{5y_{1,0}}%
\right\vert y_{1,0},y_{2,0}\right) =P\left( \left. \alpha =\sqrt{5y_{1,0}}%
\right\vert y_{1,0},y_{2,0}\right) =\frac{1}{2}$.
\end{itemize}

The results are in Table \ref{TABLE: plim of CRE}.

\begin{table}[tbp]
\caption{Probability Limit of Conditional Random Effects Estimator under
Different Heterogeneity Distributions}
\label{TABLE: plim of CRE}
\begin{center}
\begin{tabular}{lrrrrrr}
\hline
Distribution of heterogeneity & $\gamma _{11}$ & $\gamma _{12}$ & $\gamma
_{21}$ & $\gamma _{22}$ & $\rho $ & $\kappa $ \\
Correctly specified & {$2.50$} & {$-1.50$} & {$-1.50$} & {$2.50$} & {$1.00$}
& {$2.00$} \\
Discrete, but approximately normal & {$2.51$} & {$-1.50$} & {$-1.52$} & {$%
2.49$} & {$0.99$} & {$2.02$} \\
Discrete, asymmetric & {$2.66$} & {$-1.73$} & {$-1.61$} & {$2.42$} & {$0.95$}
& {$2.08$} \\
Heteroskedastic & {$2.68$} & {$-1.62$} & {$-1.92$} & {$2.44$} & {$1.00$} & {$%
2.39$} \\
Very heteroskedastic & {$2.64$} & {$-1.29 $} & {$-1.91$} & {$2.63$} & {$1.27$%
} & {$2.53$}%
\end{tabular}%
\end{center}
\par
{\footnotesize The table gives the probability limit of the correlated
random effects estimator for various distributions of the fixed effect when $%
\left( \gamma _{11},\gamma _{12},\gamma _{21},\gamma _{22},\rho ,\kappa
\right) =\left( 2.5,-1.5,-1.5,2.5,1,2\right) $ and $y_{1,0}$ and $y_{2,0}$
are independent and equal to 1 with probability $\frac{1}{2}$.}
\end{table}

The probability limits in Table \ref{TABLE: plim of CRE} illustrate that the
correlated random effects approach can provide a very good approximation
when the distribution of the heterogeneity ($\alpha$) is well-approximated
by the assumed functional form, but also that the biases can be a much
larger source of estimation error for the estimator than sampling variance
for the kind of sample sizes considered here.

\section{Conclusion}

Two of Peter Schmidt's many contributions to econometrics have been to
introduce an econometric model for simultaneous binary outcomes and to study
the estimation of dynamic linear fixed effects panel data models using short
panels. In this paper, we combine aspects of this research by studying %
panel data versions of the model introduced in \cite%
{SchmidtStrauss1975} that allow for lagged dependent variables and fixed
effects,
and we apply existing as well as new methods to investigate
the joint behavior of employment of husbands and wives.

On the methodological side, we first use the conditional likelihood approach of \cite{HonoreKyriazidou2019a} to construct a likelihood function that does not
depend on the fixed effects of the model. While this conditional likelihood
can be used to estimate the other parameters of the model when the total number of time periods is at least four, it turns out that it does not depend on
the parameter $\rho$, which in the Schmidt-Strauss model captures the
inter-equation dependence. As a result, our conditional likelihood approach can
 not be used to estimate this parameter.
We therefore next use the approach in \cite{HonoreWeidner2020} to study whether one
can construct moment conditions that can be used to estimate $\rho$. We find
that it is in principle possible to estimate the common parameters of such
models when the total number of time periods for each individual is at least
five. To construct moment conditions for four time periods, it is necessary
to restrict the model. We do this by restricting the fixed effects for the
two outcomes to be equal, except for an additive constant.

On the empirical side,  we apply existing methods like those developed in \cite{SchmidtStrauss1975}, as well as  the estimation methods developed in this paper, to estimate a
simple model for the relationship of employment of husbands and wives. Our
main conclusion is that the parameter that captures the intra-household
dependence in employment varies by the ethnicity composition of the
couple and  over time, even after one allows for unobserved household specific heterogeneity.

\ifx\undefined\BySame
\newcommand{\BySame}{\leavevmode\rule[.5ex]{3em}{.5pt}\ }
\fi
\ifx\undefined\textsc
\newcommand{\textsc}[1]{{\sc #1}}
\newcommand{\emph}[1]{{\em #1\/}}
\let\tmpsmall\small
\renewcommand{\small}{\tmpsmall\sc}
\fi

\section*{Appendix: Moment Conditions}

In this Appendix, we explicitly present the six moment conditions discussed
in Section \ref{Some Moment Conditions}. To simplify the notation, we write $%
\Gamma _{ij}=\exp \left( \gamma _{ij}\right) $, $B=\exp \left( \beta \right)
$, and $P=\exp \left( \rho \right) $.

\subsection*{Moment Condition 1}

\begin{equation*}
E\left[ \sum\limits_{k=1}^{5}m_{k}1\left\{ \left( \left\{ y_{1,t}\right\}
_{t=0}^{3},\left\{ y_{2,t}\right\} _{t=0}^{3}\right) =s_{k}\right\} \right]
=0
\end{equation*}%
where%
\begin{eqnarray*}
s_{1} &=&(a,0,0,1,b,0,1,0) \\
s_{2} &=&(a,0,0,1,b,0,1,1) \\
s_{3} &=&(a,0,1,0,b,0,1,0) \\
s_{4} &=&(a,0,1,0,b,1,1,0) \\
s_{5} &=&(a,0,1,1,b,0,1,0)
\end{eqnarray*}%
and
\begin{multline*}
m_{1}=B\Gamma _{11}\Gamma _{22}P\Bigl[\Gamma _{12}\left( -B\Gamma
_{21}\Gamma _{22}^{2}+(B+1)\Gamma _{22}+\Gamma _{11}\left( \Gamma
_{21}\Gamma _{22}\left( B\Gamma _{22}-B-1\right) +1\right) -1\right) \Bigr.
\\
+\Bigl.B\left( \Gamma _{21}-1\right) \Gamma _{22}+\Gamma _{11}\left( \Gamma
_{21}-1\right) \Gamma _{22}\Gamma _{12}^{2}\Bigr]
\end{multline*}%
\begin{multline*}
m_{2}=\Gamma _{11}\Bigl[B^{2}\left( \Gamma _{21}-1\right) \Gamma _{22}^{2}+
\\
\Gamma _{12}^{2}\left( -B\Gamma _{22}^{2}P+\Gamma _{11}\left( B\Gamma
_{21}\Gamma _{22}^{2}P-(B+1)\Gamma _{22}+1\right) +(B+1)\Gamma
_{22}-1\right) \Bigr. \\
+\Bigl.B\Gamma _{22}\Gamma _{12}\left( B\Gamma _{22}-\Gamma _{21}\left(
(B+1)\Gamma _{22}-1\right) +\Gamma _{11}\left( 1-\Gamma _{21}P\right)
+\Gamma _{22}+P-2\right) \Bigr]
\end{multline*}%
\begin{multline*}
m_{3}=-B\Gamma _{11}\Gamma _{12}\Gamma _{22} \\
\Bigl[\Gamma _{12}\left( -B\Gamma _{21}\Gamma _{22}^{2}+(B+1)\Gamma
_{22}+\Gamma _{11}\left( \Gamma _{21}\Gamma _{22}\left( B\Gamma
_{22}-B-1\right) +1\right) -1\right) \Bigr. \\
+\Bigl.B\left( \Gamma _{21}-1\right) \Gamma _{22}+\Gamma _{11}\left( \Gamma
_{21}-1\right) \Gamma _{22}\Gamma _{12}^{2}\Bigr]
\end{multline*}%
\begin{multline*}
m_{4}=-\Gamma _{11}\Gamma _{12}\Gamma _{21}^{-a}\Gamma _{22}^{1-b}\Bigl[%
\Gamma _{11} \\
\left( B^{2}\left( \Gamma _{21}-1\right) \Gamma _{21}\Gamma
_{22}^{2}+B\Gamma _{22}\Gamma _{12}\left( 2\Gamma _{21}(P-2)+\Gamma
_{21}^{2}+1\right) -\left( \left( \Gamma _{21}-1\right) \Gamma
_{12}^{2}\right) \right) \Bigr. \\
+\Bigl.\Gamma _{12}\Gamma _{11}^{2}\left( B\Gamma _{21}\Gamma _{22}\left(
1-\Gamma _{21}P\right) +\Gamma _{12}\left( \Gamma _{21}-1\right) \right)
+B\Gamma _{22}\left( \Gamma _{12}\left( \Gamma _{21}-P\right) -B\left(
\Gamma _{21}-1\right) \Gamma _{21}\Gamma _{22}\right) \Bigr]
\end{multline*}%
\begin{multline*}
m_{5}=B\Gamma _{22}\Bigl[\Gamma _{11}^{2}\Gamma _{12}^{2}\left( -B\Gamma
_{21}^{2}\Gamma _{22}^{2}P+(B+1)\Gamma _{21}\Gamma _{22}-1\right) \Bigr. \\
+\Gamma _{11}\Gamma _{12}\bigl( B\Gamma _{22}\left( \Gamma _{21}\left(
-(B+1)\Gamma _{22}+P-2\right) +(B+1)\Gamma _{22}\Gamma _{21}^{2}+1\right) \\
+\Gamma _{12}\left( \Gamma _{21}\Gamma _{22}\left( B\Gamma _{22}P-B-1\right)
+1\right) \bigr) \\
+\Bigl.B\Gamma _{22}\left( \Gamma _{12}\left( \Gamma _{21}-P\right) -B\left(
\Gamma _{21}-1\right) \Gamma _{21}\Gamma _{22}\right) \Bigr]
\end{multline*}

\subsection*{Moment Condition 2}

\begin{equation*}
E\left[ \sum\limits_{k=1}^{5}m_{k}1\left\{ \left( \left\{ y_{1,t}\right\}
_{t=0}^{3},\left\{ y_{2,t}\right\} _{t=0}^{3}\right) =s_{k}\right\} \right]
=0
\end{equation*}%
where%
\begin{eqnarray*}
s_{1} &=&(a,0,0,1,b,1,0,0) \\
s_{2} &=&(a,0,0,1,b,1,0,1) \\
s_{3} &=&(a,0,1,0,b,0,1,1) \\
s_{4} &=&(a,1,0,0,b,1,0,0) \\
s_{5} &=&(a,1,0,0,b,1,1,0)
\end{eqnarray*}%
and%
\begin{multline*}
m_{1}=-B\Gamma _{21}P\Gamma _{11}^{a}\Gamma _{12}^{b}\left[ \Gamma
_{12}\left( -B\Gamma _{21}\Gamma _{22}^{2}+(B+1)\Gamma _{22}+\Gamma
_{11}\left( \Gamma _{21}\Gamma _{22}\left( B\Gamma _{22}-B-1\right)
+1\right) -1\right) \right. \\
+\left. B\left( \Gamma _{21}-1\right) \Gamma _{22}+\Gamma _{11}\left( \Gamma
_{21}-1\right) \Gamma _{22}\Gamma _{12}^{2}\right]
\end{multline*}%
\begin{multline*}
m_{2}=\Gamma _{21}\left( -\Gamma _{11}^{a}\right) \Gamma _{12}^{b}\left[
B^{2}\left( \Gamma _{21}-1\right) \Gamma _{22}^{2}\right. \\
+\Gamma _{12}^{2}\left( -B\Gamma _{22}^{2}P+\Gamma _{11}\left( B\Gamma
_{21}\Gamma _{22}^{2}P-(B+1)\Gamma _{22}+1\right) +(B+1)\Gamma _{22}-1\right)
\\
+\left. B\Gamma _{22}\Gamma _{12}\left( B\Gamma _{22}-\Gamma _{21}\left(
(B+1)\Gamma _{22}-1\right) +\Gamma _{11}\left( 1-\Gamma _{21}P\right)
+\Gamma _{22}+P-2\right) \right]
\end{multline*}%
\begin{multline*}
m_{3}=\Gamma _{11}^{a}\Gamma _{21}^{a}\Gamma _{12}^{b}\Gamma _{22}^{b-1}
\left[ \Gamma _{11}^{2}\Gamma _{12}^{2}\left( B\Gamma _{21}^{2}\Gamma
_{22}^{2}P-(B+1)\Gamma _{21}\Gamma _{22}+1\right) \right. \\
-\Gamma _{11}\Gamma _{12}\bigl( \ B\Gamma _{22}\left( \Gamma _{21}\left(
-(B+1)\Gamma _{22}+P-2\right) +(B+1)\Gamma _{22}\Gamma _{21}^{2}+1\right) \\
+\Gamma _{12}\left( \Gamma _{21}\Gamma _{22}\left( B\Gamma _{22}P-B-1\right)
+1\right) \bigr) \\
+\left. B\Gamma _{22}\left( B\left( \Gamma _{21}-1\right) \Gamma _{21}\Gamma
_{22}+\Gamma _{12}\left( P-\Gamma _{21}\right) \right) \right]
\end{multline*}%
\begin{multline*}
m_{4}=B\Gamma _{21}\left[ \Gamma _{12}\left( -B\Gamma _{21}\Gamma
_{22}^{2}+(B+1)\Gamma _{22}+\Gamma _{11}\left( \Gamma _{21}\Gamma
_{22}\left( B\Gamma _{22}-B-1\right) +1\right) -1\right) \right. \\
+\left. B\left( \Gamma _{21}-1\right) \Gamma _{22}+\Gamma _{11}\left( \Gamma
_{21}-1\right) \Gamma _{22}\Gamma _{12}^{2}\right]
\end{multline*}%
\begin{multline*}
m_{5}=\Gamma _{12}\left[ \Gamma _{11}\left( B^{2}\left( \Gamma
_{21}-1\right) \Gamma _{21}\Gamma _{22}^{2}+B\Gamma _{22}\Gamma _{12}\left(
2\Gamma _{21}(P-2)+\Gamma _{21}^{2}+1\right) -\left( \left( \Gamma
_{21}-1\right) \Gamma _{12}^{2}\right) \right) \right. + \\
\left. \Gamma _{12}\Gamma _{11}^{2}\left( B\Gamma _{21}\Gamma _{22}\left(
1-\Gamma _{21}P\right) +\Gamma _{12}\left( \Gamma _{21}-1\right) \right)
+B\Gamma _{22}\left( \Gamma _{12}\left( \Gamma _{21}-P\right) -B\left(
\Gamma _{21}-1\right) \Gamma _{21}\Gamma _{22}\right) \right]
\end{multline*}

\subsection*{Moment Condition 3}

\begin{equation*}
E\left[ \sum\limits_{k=1}^{5}m_{k}1\left\{ \left( \left\{ y_{1,t}\right\}
_{t=0}^{3},\left\{ y_{2,t}\right\} _{t=0}^{3}\right) =s_{k}\right\} \right]
=0
\end{equation*}%
where%
\begin{eqnarray*}
s_{1} &=&(a,0,1,1,b,1,0,1) \\
s_{2} &=&(a,0,1,1,b,1,1,1) \\
s_{3} &=&(a,1,0,0,b,1,1,1) \\
s_{4} &=&(a,1,1,1,b,0,1,0) \\
s_{5} &=&(a,1,1,1,b,0,1,1)
\end{eqnarray*}%
and%
\begin{multline*}
m_{1}=-B\Gamma _{11}^{a+1}\Gamma _{21}^{-a}\Gamma _{12}^{b}\Gamma _{22}^{1-b}
\\
\left[ \Gamma _{11}(B^{2}\Gamma _{21}\left( \Gamma _{21}-\Gamma _{22}\right)
\Gamma _{22}\right. \\
+B\Gamma _{12}\left( 2\Gamma _{22}\Gamma _{21}(P-2)+\Gamma _{21}^{2}+\Gamma
_{22}^{2}\right) +\Gamma _{12}^{2}\left( \Gamma _{22}-\Gamma _{21}\right) )
\\
+\left. \Gamma _{11}^{2}\left( B\Gamma _{21}\left( \Gamma _{22}-\Gamma
_{21}P\right) +\Gamma _{12}\left( \Gamma _{21}-\Gamma _{22}\right) \right)
+B\Gamma _{12}\Gamma _{22}\left( B\Gamma _{21}\left( \Gamma _{22}-\Gamma
_{21}\right) +\Gamma _{12}\left( \Gamma _{21}-\Gamma _{22}P\right) \right)
\right]
\end{multline*}%
\begin{multline*}
m_{2}=-B\Gamma _{11}^{a+1}\Gamma _{21}^{-a}\Gamma _{12}^{b}\Gamma _{22}^{-b}
\\
\left[ \Gamma _{11}\left( \left( \Gamma _{21}-\Gamma _{22}\right) \Gamma
_{12}\left( B\Gamma _{21}\Gamma _{22}+1\right) -B\left( \Gamma
_{21}-1\right) \Gamma _{21}\Gamma _{22}-\left( \left( \Gamma _{21}-1\right)
\Gamma _{22}\Gamma _{12}^{2}\right) \right) \right. \\
+\left. B\Gamma _{12}\Gamma _{21}\left( \Gamma _{22}-1\right) \Gamma
_{22}+\Gamma _{12}\Gamma _{21}\left( \Gamma _{22}-1\right) \Gamma _{11}^{2}
\right]
\end{multline*}%
\begin{multline*}
m_{3}=\Gamma _{11}^{2}\Gamma _{12}\Gamma _{21}^{-a}\Gamma _{22}^{-b-1} \\
\bigl[ \Gamma _{11}(B^{2}\left( \Gamma _{21}-1\right) \Gamma _{21}\Gamma
_{22}^{2}+\Gamma _{12}^{2}\left( \Gamma _{21}\left( B\Gamma
_{22}^{2}P-B\Gamma _{22}-1\right) +\Gamma _{22}\right) \\
+B\Gamma _{22}\Gamma _{12}\left( \Gamma _{21}\left( -\Gamma _{22}+P-2\right)
+\Gamma _{21}^{2}+\Gamma _{22}\right) ) \\
+\Gamma _{12}\Gamma _{11}^{2}\left( B\Gamma _{21}\Gamma _{22}\left( 1-\Gamma
_{21}P\right) +\Gamma _{12}\left( \Gamma _{21}-\Gamma _{22}\right) \right) \\
+B\Gamma _{12}\Gamma _{22}\left( \Gamma _{12}\left( \Gamma _{21}-\Gamma
_{22}P\right) -B\left( \Gamma _{21}-1\right) \Gamma _{21}\Gamma _{22}\right) %
\bigr]
\end{multline*}%
\begin{multline*}
m_{4}=B^{2}\Gamma _{22}\bigl[ B^{2}\Gamma _{12}\Gamma _{21}^{2}\left( \Gamma
_{22}-1\right) \Gamma _{22}+ \\
\Gamma _{11}^{2}\left( \Gamma _{12}\left( B\Gamma _{22}\Gamma
_{21}^{2}P+\Gamma _{21}\left( 1-B\Gamma _{22}\right) -\Gamma _{22}\right)
+B\Gamma _{21}\left( \Gamma _{22}-\Gamma _{21}P\right) +\left( \Gamma
_{22}-\Gamma _{21}\right) \Gamma _{12}^{2}\right) \\
+B\Gamma _{21}\Gamma _{11}\left( -B\Gamma _{21}\left( \Gamma _{22}-1\right)
\Gamma _{22}+\Gamma _{12}^{2}\left( \Gamma _{22}-\Gamma _{22}^{2}P\right)
+\Gamma _{12}\left( \Gamma _{22}\left( \Gamma _{22}+P-2\right) -\Gamma
_{21}\left( \Gamma _{22}-1\right) \right) \right) \bigr]
\end{multline*}%
\begin{multline*}
m_{5}=B^{2}\Gamma _{11}\left[ \Gamma _{11}\left( \left( \Gamma _{21}-\Gamma
_{22}\right) \Gamma _{12}\left( B\Gamma _{21}\Gamma _{22}+1\right) -B\left(
\Gamma _{21}-1\right) \Gamma _{21}\Gamma _{22}-\left( \left( \Gamma
_{21}-1\right) \Gamma _{22}\Gamma _{12}^{2}\right) \right) \right. \\
+\left. B\Gamma _{12}\Gamma _{21}\left( \Gamma _{22}-1\right) \Gamma
_{22}+\Gamma _{12}\Gamma _{21}\left( \Gamma _{22}-1\right) \Gamma _{11}^{2}
\right]
\end{multline*}

\subsection*{Moment Condition 4}

\begin{equation*}
E\left[ \sum\limits_{k=1}^{5}m_{k}1\left\{ \left( \left\{ y_{1,t}\right\}
_{t=0}^{3},\left\{ y_{2,t}\right\} _{t=0}^{3}\right) =s_{k}\right\} \right]
=0
\end{equation*}%
where%
\begin{eqnarray*}
s_{1} &=&(a,1,0,0,b,1,0,1) \\
s_{2} &=&(a,1,0,1,b,0,0,1) \\
s_{3} &=&(a,1,0,1,b,1,0,1) \\
s_{4} &=&(a,1,1,0,b,1,0,0) \\
s_{5} &=&(a,1,1,0,b,1,0,1)
\end{eqnarray*}%
and
\begin{multline*}
m_{1}=\Gamma _{12} \\
\left[ \Gamma _{11}\left( -\Gamma _{21}\left( B^{2}+\Gamma _{12}\left(
B-B\Gamma _{22}P\right) +B\Gamma _{22}-BP+2B+1\right) +B(B+1)\Gamma
_{22}\Gamma _{21}^{2}+B+1\right) \right. \\
+\left. \Gamma _{12}\Gamma _{11}^{2}\left( -B\Gamma _{22}\Gamma
_{21}^{2}P+(B+1)\Gamma _{21}-1\right) +B\left( -B\Gamma _{22}\Gamma
_{21}^{2}+(B+1)\Gamma _{21}-P\right) \right]
\end{multline*}%
\begin{multline*}
m_{2}=-B\Gamma _{12}\Gamma _{21}^{a}\Gamma _{22}^{b}\left[ \Gamma
_{11}\left( -\Gamma _{21}\left( B^{2}-2BP+4B+1\right) +B(B+1)\Gamma
_{21}^{2}+B+1\right) \right. \\
+\left. \Gamma _{11}^{2}\left( -B\Gamma _{21}^{2}P+(B+1)\Gamma
_{21}-1\right) +B\left( -B\Gamma _{21}^{2}+(B+1)\Gamma _{21}-P\right) \right]
\end{multline*}%
\begin{multline*}
m_{3}=-B\Gamma _{12}\left[ \Gamma _{11}\left( -B\Gamma _{22}\Gamma
_{21}^{2}+(B+1)\Gamma _{21}+\Gamma _{12}\left( B\Gamma _{22}\Gamma
_{21}^{2}-(B+1)\Gamma _{22}\Gamma _{21}+1\right) -1\right) \right. \\
+\left. B\Gamma _{21}\left( \Gamma _{22}-1\right) +\Gamma _{12}\Gamma
_{21}\left( \Gamma _{22}-1\right) \Gamma _{11}^{2}\right]
\end{multline*}%
\begin{multline*}
m_{4}=B\left[ B^{2}\left( \Gamma _{22}-1\right) \Gamma _{21}^{2}\Gamma
_{11}^{-1}\right. \\
+B\Gamma _{21}\left( -(B+1)\Gamma _{21}\left( \Gamma _{22}-1\right) +\Gamma
_{12}\left( 1-\Gamma _{22}P\right) +\Gamma _{22}+P-2\right) \\
+\left. \Gamma _{11}\left( -B\Gamma _{21}^{2}P+\Gamma _{12}\left( B\Gamma
_{22}\Gamma _{21}^{2}P-(B+1)\Gamma _{21}+1\right) +(B+1)\Gamma
_{21}-1\right) \right]
\end{multline*}%
\begin{multline*}
m_{5}=BP\left[ \Gamma _{11}\left( -B\Gamma _{22}\Gamma _{21}^{2}+(B+1)\Gamma
_{21}+\Gamma _{12}\left( B\Gamma _{22}\Gamma _{21}^{2}-(B+1)\Gamma
_{22}\Gamma _{21}+1\right) -1\right) \right. \\
+\left. B\Gamma _{21}\left( \Gamma _{22}-1\right) +\Gamma _{12}\Gamma
_{21}\left( \Gamma _{22}-1\right) \Gamma _{11}^{2}\right]
\end{multline*}

\subsection*{Moment Condition 5}

\begin{equation*}
E\left[ \sum\limits_{k=1}^{5}m_{k}1\left\{ \left( \left\{ y_{1,t}\right\}
_{t=0}^{3},\left\{ y_{2,t}\right\} _{t=0}^{3}\right) =s_{k}\right\} \right]
=0
\end{equation*}%
where%
\begin{eqnarray*}
s_{1} &=&(a,0,0,0,b,0,1,0) \\
s_{2} &=&(a,0,0,0,b,0,1,1) \\
s_{3} &=&(a,0,1,0,b,0,0,0) \\
s_{4} &=&(a,0,1,0,b,0,0,1) \\
s_{5} &=&(a,0,1,0,b,1,0,0)
\end{eqnarray*}%
and
\begin{multline*}
m_{1}=B\Gamma _{12}\Gamma _{21}^{a}\Gamma _{22}^{b}\left[ B\left( \Gamma
_{22}-\Gamma _{21}\right) +\Gamma _{12}\left( \Gamma _{22}\left( B\Gamma
_{21}-B-1\right) +1\right) \right. \\
+\left. \Gamma _{11}\left( \Gamma _{21}\left( B\left( -\Gamma _{22}\right)
+B-\Gamma _{12}+1\right) +\Gamma _{12}\Gamma _{22}-1\right) \right]
\end{multline*}%
\begin{multline*}
m_{2}=\Gamma _{12}\Gamma _{21}^{a}\Gamma _{22}^{b-1} \\
\left[ B^{2}\Gamma _{22}\left( \Gamma _{22}-\Gamma _{21}\right) +B\Gamma
_{12}\left( \Gamma _{21}\left( (B+1)\Gamma _{22}-1\right) -\Gamma
_{22}\left( (B+1)\Gamma _{22}+P-2\right) \right) \right. \\
+\Gamma _{12}^{2}\left( B\Gamma _{22}^{2}P-(B+1)\Gamma _{22}+1\right) \\
\left. +\Gamma _{11}\left( B\left( \Gamma _{21}P-\Gamma _{22}\right) +\Gamma
_{12}\left( \Gamma _{22}\left( B\Gamma _{21}(-P)+B+1\right) -1\right)
\right) \right]
\end{multline*}%
\begin{multline*}
m_{3}=-B^{2}\Gamma _{12}\Gamma _{21}^{a}\Gamma _{22}^{b}\left[ B\left(
\Gamma _{22}-\Gamma _{21}\right) +\Gamma _{12}\left( \Gamma _{22}\left(
B\Gamma _{21}-B-1\right) +1\right) \right. \\
+\left. \Gamma _{11}\left( \Gamma _{21}\left( B\left( -\Gamma _{22}\right)
+B-\Gamma _{12}+1\right) +\Gamma _{12}\Gamma _{22}-1\right) \right]
\end{multline*}%
\begin{multline*}
m_{4}=-B\Gamma _{12}\Gamma _{21}^{a-1}\Gamma _{22}^{b}\left[ \Gamma
_{11}^{2}\left( -B\Gamma _{21}^{2}P+(B+1)\Gamma _{21}-1\right) \right. \\
+\Gamma _{11}\bigl[ B\left( \Gamma _{21}\left( -(B+1)\Gamma _{22}+P-2\right)
+(B+1)\Gamma _{21}^{2}+\Gamma _{22}\right) \\
+\Gamma _{12}\left( \Gamma _{21}\left( B\Gamma _{22}P-B-1\right) +1\right) %
\bigr] \\
+\left. B\left( B\Gamma _{21}\left( \Gamma _{22}-\Gamma _{21}\right) +\Gamma
_{12}\left( \Gamma _{21}-\Gamma _{22}P\right) \right) \right]
\end{multline*}%
\begin{multline*}
m_{5}=B\left[ \Gamma _{11}\left( B^{2}\Gamma _{21}\left( \Gamma _{21}-\Gamma
_{22}\right) \Gamma _{22}+B\Gamma _{12}\left( 2\Gamma _{22}\Gamma
_{21}(P-2)+\Gamma _{21}^{2}+\Gamma _{22}^{2}\right) +\Gamma _{12}^{2}\left(
\Gamma _{22}-\Gamma _{21}\right) \right) \right. \\
+\left. \Gamma _{11}^{2}\left( B\Gamma _{21}\left( \Gamma _{22}-\Gamma
_{21}P\right) +\Gamma _{12}\left( \Gamma _{21}-\Gamma _{22}\right) \right)
+B\Gamma _{12}\Gamma _{22}\left( B\Gamma _{21}\left( \Gamma _{22}-\Gamma
_{21}\right) +\Gamma _{12}\left( \Gamma _{21}-\Gamma _{22}P\right) \right)
\right]
\end{multline*}

\subsection*{Moment Condition 6}

\begin{equation*}
E\left[ \sum\limits_{k=1}^{5}m_{k}1\left\{ \left( \left\{ y_{1,t}\right\}
_{t=0}^{3},\left\{ y_{2,t}\right\} _{t=0}^{3}\right) =s_{k}\right\} \right]
=0
\end{equation*}%
where%
\begin{eqnarray*}
s_{1} &=&(a,0,0,0,b,1,0,1) \\
s_{2} &=&(a,0,1,0,b,1,0,1) \\
s_{3} &=&(a,0,1,1,b,0,0,0) \\
s_{4} &=&(a,0,1,1,b,0,0,1) \\
s_{5} &=&(a,1,0,0,b,0,1,0)
\end{eqnarray*}%
and
\begin{multline*}
m_{1}=\Gamma _{11}^{a}\Gamma _{21}^{1-a}\Gamma _{12}^{b}\Gamma _{22}^{-b}
\left[ -\Gamma _{12}\left( -\Gamma _{22}\left( B^{2}-2BP+4B+1\right)
+B(B+1)\Gamma _{22}^{2}+B+1\right) \right. \\
+\left. \Gamma _{12}^{2}\left( B\Gamma _{22}^{2}P-(B+1)\Gamma _{22}+1\right)
+B\left( B\Gamma _{22}^{2}-(B+1)\Gamma _{22}+P\right) \right]
\end{multline*}%
\begin{multline*}
m_{2}=BP\Gamma _{11}^{a}\Gamma _{21}^{-a}\Gamma _{12}^{b}\Gamma _{22}^{1-b}
\left[ B\left( \Gamma _{21}-\Gamma _{22}\right) +\Gamma _{12}\left( \Gamma
_{22}\left( B\left( -\Gamma _{21}\right) +B+1\right) -1\right) \right. \\
+\left. \Gamma _{11}\left( \Gamma _{21}\left( B\Gamma _{22}-B+\Gamma
_{12}-1\right) -\Gamma _{12}\Gamma _{22}+1\right) \right]
\end{multline*}%
\begin{multline*}
m_{3}=B^{2}\Gamma _{21}\Gamma _{11}^{a-1}\Gamma _{12}^{b} \\
\bigl[ \Gamma _{12}(-B^{2}\Gamma _{22}+B\Gamma _{22}P+\Gamma _{11}\left(
\Gamma _{22}\left( B\Gamma _{21}(-P)+B+1\right) -1\right) \\
-2B\Gamma _{22}+B\Gamma _{21}\left( (B+1)\Gamma _{22}-1\right) +B-\Gamma
_{22}+1) \\
+B\left( \Gamma _{22}\left( B\left( -\Gamma _{21}\right) +B+1\right) +\Gamma
_{11}\left( \Gamma _{21}P-\Gamma _{22}\right) -P\right) \bigr]
\end{multline*}%
\begin{eqnarray*}
m_{4} &=&B^{2}\Gamma _{11}^{a-1}\Gamma _{12}^{b}\left[ B\left( \Gamma
_{22}-\Gamma _{21}\right) +\Gamma _{12}\left( \Gamma _{22}\left( B\Gamma
_{21}-B-1\right) +1\right) \right. \\
&&+\left. \Gamma _{11}\left( \Gamma _{21}\left( B\left( -\Gamma _{22}\right)
+B-\Gamma _{12}+1\right) +\Gamma _{12}\Gamma _{22}-1\right) \right]
\end{eqnarray*}%
\begin{multline*}
m_{5}=B\left[ B^{2}\left( \Gamma _{21}-\Gamma _{22}\right) \Gamma
_{22}+B\Gamma _{12}\left( \Gamma _{22}\left( (B+1)\Gamma _{22}+P-2\right)
-\Gamma _{21}\left( (B+1)\Gamma _{22}-1\right) \right) \right. \\
+\Gamma _{12}^{2}\left( -B\Gamma _{22}^{2}P+(B+1)\Gamma _{22}-1\right) \\
+\left. \Gamma _{11}\left( B\left( \Gamma _{22}-\Gamma _{21}P\right) +\Gamma
_{12}\left( \Gamma _{22}\left( B\Gamma _{21}P-B-1\right) +1\right) \right)
\right]
\end{multline*}

\end{document}